\documentclass[lettersize,journal]{IEEEtran}
\usepackage{amsmath,amsfonts}
\usepackage{algorithmic}
\usepackage{algorithm}
\usepackage{array}
\usepackage[caption=false,font=normalsize,labelfont=sf,textfont=sf]{subfig}
\usepackage{textcomp}
\usepackage{stfloats}
\usepackage{url}
\usepackage{verbatim}
\usepackage{graphicx}
\usepackage{cite}
\usepackage{multirow}
\usepackage{booktabs}
\usepackage{amssymb}

\usepackage{hyperref}

\begin{document}

\title{SleepBand: Single-Source Domain Generalization for Sleep Staging via Physiologically Structured Spectral Modeling}

\author{Zhi Lu
        Yang Hu,~\IEEEmembership{Member,~IEEE}, 
        and Yan Chen,~\IEEEmembership{Senior Member,~IEEE}%
\thanks{Zhi Lu is with the Laboratory of Intelligent Collaborative Computing, University of Electronic Science and Technology of China, Chengdu 611731, China. (e-mail: zhilu@uestc.edu.cn)}
\thanks{Yang Hu is with the School of Information Science and Technology, University of Science and Technology of China, Hefei 230026, China. (e-mail: eeyhu@ustc.edu.cn)}
\thanks{Yan Chen is with the School of Cyber Science and Technology, University of Science and Technology of China, Hefei 230026, China. (e-mail: eecyan@ustc.edu.cn)}%
\thanks{Corresponding Author: Yan Chen}
}

\markboth{Journal of \LaTeX\ Class Files,~Vol.~14, No.~8, August~2021}%
{Shell \MakeLowercase{\textit{et al.}}: A Sample Article Using IEEEtran.cls for IEEE Journals}


\maketitle

\begin{abstract}
Generalizing sleep staging models to unseen datasets is challenging, and typical domain generalization (DG) methods often rely on multiple source domains or domain labels that are rarely available in practice. We tackle the stricter and more practical setting of single-source domain generalization: training on a single labeled source dataset, without domain labels or access to target data. We present SleepBand, a physiology-guided framework that embeds oscillatory priors via a learnable Morlet filter bank and a structured integration-and-recalibration pipeline. This anchors representations to domain-invariant sleep rhythms (e.g., slow waves, spindles), reducing reliance on dataset-specific artefacts. On five public datasets, SleepBand achieves state-of-the-art SDG performance and remains competitive under leave-one-domain-out (multi-source) DG. Analyses show that the learned filters align with canonical neurophysiology and that robustness stems from focusing on narrowband, physiologically meaningful cues. Our results suggest that principled, physiology-aware inductive biases are a promising path for robust single-domain sleep staging. Code is available at~\url{https://github.com/lzcn/sleep-band}.
\end{abstract}


\section{Introduction}
\IEEEPARstart{S}{leep} staging is the task to assign each 30-second epoch of polysomnographic (PSG) recordings to one of the canonical sleep stages (W, N1, N2, N3, or REM) and serves as a cornerstone of sleep health assessment~\cite{SleepHeartHealthQuan97,TransferableSelfSupervisedInstanceZhao22,MultiviewMultimodalSystemTorres18}. While recent deep learning models have achieved strong in-distribution performance~\cite{DeepSleepNetModelAutomaticSupratak17}, they often suffer significant degradation on unseen datasets due to distribution shifts caused by variations in hardware, montage, preprocessing, and subject populations~\cite{ATTAAdaptiveTestTimeJia24}. To mitigate such shifts, domain generalization (DG) methods aim to learn domain-invariant representations, typically by aligning features across multiple source domains~\cite{DeepCORALCorrelationSun16,DomainGeneralizationAdversarialLi18}. In practice, however, curating diverse clinical datasets that capture the full spectrum of real-world variability is challenging, motivating the study of single-source domain generalization (SDG)~\cite{ProgressiveDomainExpansionLi21,PracticalSingleDomainYang24,LearningDiversifySingleWang21,ProgressiveInvariantCausalWang25,LearningLearnSingleQiao20}. 
Existing cross-domain efforts focus primarily on subject-level adaptation~\cite{MultiViewSpatialTemporalGraphJia21,DomainInvariantRepresentationLearningLee25} or assume access to multiple source domains~\cite{GeneralizableSleepStagingWang24}, leaving a critical gap for robust single-source generalization.


\begin{figure}[!t]
    \centering
    \includegraphics[width=0.98\linewidth]{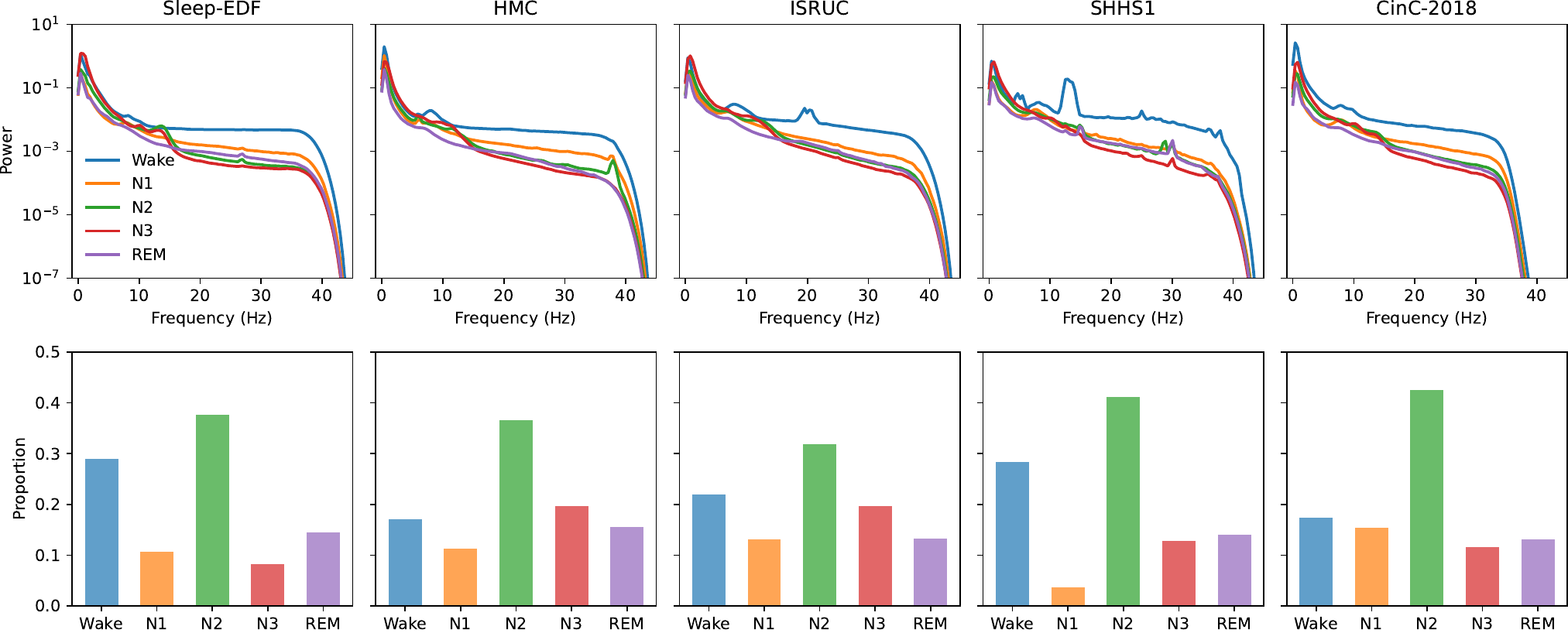}
\caption{Comparison of EEG spectral profiles and sleep stage distributions across datasets. For each dataset, stage-wise power spectral densities (top) and normalized stage proportions (bottom) are presented. Although spectral trends remain broadly consistent, the distribution of sleep stages varies significantly, highlighting potential domain shifts and dataset heterogeneity.}
    \label{fig:stage_distribution}
\end{figure}

To demonstrate the ubiquity of distribution shifts and uniqueness in sleep staging, we show the signal characteristics and label distributions of data collected from different environments in Fig.~\ref{fig:stage_distribution}. Although these data exhibit similar overall statistical characteristics, noticeable differences still persist.  
For example, the EEG spectral profiles from the SHHS1 and ISRUC datasets show more pronounced artifacts during weakness. Additionally, although not a clinical diagnostic criterion, the relative energy ordering across sleep stages differs, with the Wake stage in the CinC 2018 dataset exhibiting higher spectral power in the low-frequency bands compared to the other datasets. These patterns, some of which are stage-related, are typically unidentifiable without prior physiological knowledge. Consequently, they can lead to spurious causal dependencies during training or cause models to misgeneralize when acted upon as out-of-distribution data at test time.


Unlike general semantic recognition tasks where objects can be characterized by diverse and redundant features, sleep staging adheres to standardized scoring criteria based on a constrained set of well-defined, narrowband oscillatory patterns (e.g., alpha rhythms for wakefulness, K‑complexes and spindles for N2 sleep)~\cite{AASMManualScoringTroester23}. These physiological oscillations are relatively stable across individuals and recording setups, yet sleep signals are also highly susceptible to non‑neural artifacts  that manifest in dataset‑specific ways. Consequently, learning within such a physiologically constrained and artifact‑prone feature space often leads models to overfit to dataset‑specific noise rather than generalizable sleep physiology.

Existing DG approaches usually learn domain-invariant representations by mitigate covariate shift $p(x)$ across multiple source domains to remove domain-specific factors, or by augmenting data when only one source is available. However, most existing works for multi-source setting often assume stable label distributions $p(x)$ or negligible concept shift $p(y|x)$ across domains~\cite{DomainGeneralizationConditionalLiu21,DomainGeneralizationSurveyZhou23}, whereas we have demonstrated their instability across datasets in the sleep staging task, as evidenced in Fig.~\ref{fig:stage_distribution}. 
In single-source settings, the constrained feature space of sleep signals makes it difficult to generate meaningful diversity and may even corrupt the physiologically meaningful patterns that define sleep stages. More critically, these approaches fail to constrain the feature space towards the underlying invariant physiological bias. This underscores the importance of incorporating physiological prior information, which is often overlooked by current approaches in sleep staging~\cite{GeneralizableSleepStagingWang24}.

To bridge this gap, we propose SleepBand, a physiology-aware framework for robust sleep staging under domain shifts. Our core design principle is to bypass the challenging explicit alignment of features across domains. Instead, SleepBand directly embeds inductive biases derived from sleep physiology to guide the model towards domain-invariant, oscillatory patterns. This is achieved through a structured encoder with three key components. A restricted learnable Gabor filterbank frontend projects input signals onto a set of narrowband, physiologically-grounded spectral kernels, anchoring feature extraction to canonical sleep rhythms such as spindles and slow waves~\cite{SleepSpindlesElectrographicCoppieterstWallant16}. A band-wise temporal integration and recalibration module models dynamics within and across frequency bands, refining the hypothesis space to emphasize robust interactions. A spectral consistency regularization strategy enforces invariance to controlled, band-wise perturbations, further steering the model away from dataset-specific spectral artefacts. By construction, SleepBand reduces reliance on spurious correlations and is applicable to both single-source and multi-source generalization settings.

Our main contributions are summarized as follows:
\begin{itemize}
	\item We formalize and address the challenging problem of single-source domain generalization for sleep staging, a realistic yet underexplored setting where models must generalize from only one labeled dataset without domain labels or target data.
	\item We introduce SleepBand, a physiology-guided framework that embeds inductive biases via a structured, learnable filter bank and adaptive spectral integration. This design actively steers feature learning toward domain-invariant sleep rhythms, reducing reliance on spurious artifacts.
	\item Extensive evaluations on five public datasets show that SleepBand achieves state-of-the-art performance in both single- and multi-source settings. Our analyses confirm that the learned representations align with canonical sleep oscillations, enhancing both robustness and interpretability.
\end{itemize}

\section{Related Work}
\subsection{Sleep Staging}
Sleep staging is typically formulated as a sequence-to-sequence classification task, extracting per-epoch representations and modeling temporal dependencies~\cite{SystematicReviewSleepWara25,WNSleepModelingWholeNightZhou25,PriorSleepNetEnhancingSleepShu25,LearningUnifiedModelYe24}.  While models, ranging from CNN-RNN hybrids~\cite{DeepSleepNetModelAutomaticSupratak17} to transformer-based architectures~\cite{SleepTransformerAutomaticSleepPhan22}, achieve strong in-distribution performance, they often generalize poorly to unseen subjects or external datasets~\cite{AttentiveAdversarialNetworkNasiri20,DeepTransferLearningRadha21,ADASTAttentiveCrossdomainEldele23,ExploringStructureIncentiveMa24}. This degradation is closely associated with inter-subject variability, such as differences in electrode placement, signal characteristics, and population-level factors, which may introduce stage-dependent, non-physiological signal components that undermine generalization.

Recently, researchers have explored transfer learning and domain adaptation to address these issues. Common strategies include pretraining on large labeled datasets followed by fine-tuning on target sets~\cite{MoreAccurateAutomaticPhan21,PersonalizationAutomaticSleepLorenzen24} or training on multiple heterogeneous datasets for robustness~\cite{RobustSleepNetTransferLearningGuillot21}. Unsupervised domain adaptation further mitigates the need for labeled target data by aligning source and unlabeled target features~\cite{DACADDomainAdaptationDarban25}, utilizing discrepancy-based~\cite{UnsupervisedDomainAdaptationFan22} or adversarial methods~\cite{UnsupervisedSleepStagingZhao21,AttentiveAdversarialNetworkNasiri20,TransferringStructuredKnowledgeYoo22,ADASTAttentiveCrossdomainEldele23}. Self-supervised learning also shows promise in learning robust features~\cite{SelfsupervisedContrastiveLearningJiang21}. 

However, a critical limitation of these methods lies in their reliance on access to the target domain.  For domain generalization, only a limited number of studies have investigated sleep staging, such as~\cite{GeneralizableSleepStagingWang24}, and within this line of work, scenarios with only a single source domain available during training, often due to privacy constraints or data acquisition costs, remain largely unexplored.

%

\subsection{Domain Generalization}
Without substantive knowledge of the target problem, no model can be expected to consistently outperform all others in domain generalization~\cite{SIMPLESpecializedModelSampleLi23}.

Domain generalization aims to train models that can generalize to unseen domains without access to target-domain data, making it a stricter and more challenging problem than domain adaptation~\cite{DomainGeneralizationSurveyZhou23}. Most existing DG studies in sleep staging adopt a multi-source setting, where data from multiple subjects or devices are used to improve cross-domain robustness. For instance, Wang et al.~\cite{GeneralizableSleepStagingWang24} proposed a multi-source DG framework that aligns representations at both the epoch and sequence levels, while Zhang et al.~\cite{SleepStagingModelZhang25} leveraged adversarial training to enhance cross-subject generalization. These methods demonstrate that DG can substantially improve model robustness in unseen recording conditions.

However, they largely rely on domain labels or multiple source datasets, which limits their applicability in single-source scenarios. In contrast, single-source domain generalization ocuses on learning domain-invariant representations from only one source domain, making it a more practical yet challenging setting for sleep staging. Despite its potential, SSDG remains underexplored in this context, and existing approaches often directly adapt generic DG techniques with limited incorporation of sleep-specific physiological knowledge~\cite{MultiViewSpatialTemporalGraphJia21,DomainInvariantRepresentationLearningLee25}. This highlights the need for SSDG strategies that are better aligned with the properties and constraints of sleep data.

However, most existing methods primarily adapt generic DG techniques with limited consideration of sleep-specific knowledge~\cite{MultiViewSpatialTemporalGraphJia21,DomainInvariantRepresentationLearningLee25}, which may constrain their performance. In addition, many approaches rely on domain labels or multiple source datasets, making them less applicable to single-source DG scenarios. These limitations suggest the need for DG strategies better suited to the characteristics of sleep staging.

Most existing single-domain generalization methods rely on data augmentation or generation. From a broader perspective, these approaches can be interpreted as bias engineering, where beneficial inductive biases such as style diversification~\cite{LearningDiversifySingleWang21,FourierbasedFrameworkDomainXu21,DomainRandomizationPyramidYue19}, structural perturbation~\cite{GeneralizingUnseenDomainsVolpi18,ProgressiveRandomConvolutionsChoi23,RobustGeneralizableVisualXu21}, or shape awareness~\cite{ReducingDomainGapNam21,ImageNettrainedCNNsAreGeirhos19} are intentionally introduced to counteract spurious biases like over-reliance on domain-specific textures or backgrounds. This perspective suggests that the design of augmentation or regularization strategies is inherently modality dependent, since different data modalities (for example, RGB, depth, sketch, or medical imaging) exhibit distinct forms of bias and therefore require different priors to guide generalization.

\begin{figure*}[!t]
    \centering
    \includegraphics[width=0.95\linewidth]{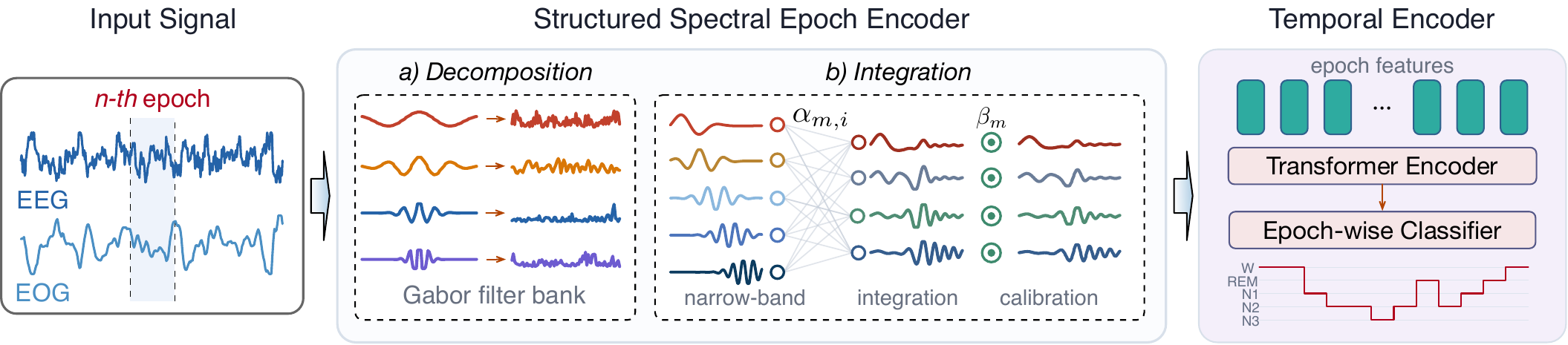}
    \caption{Overview of the proposed model architecture. For each epoch, the raw EEG/EOG signals are first fed into the structured spectral epoch encoder, where a Gabor filter bank is used for sub-band decomposition and the resulting narrow-band features are integrated to form the epoch representation. The sequence of epoch representations is then passed to the temporal encoder for cross-epoch dependency modeling, followed by epoch-wise classification.}
    \label{fig:framework}
\end{figure*}

\section{Preliminaries}
\subsection{Problem Setting}
Let $\mathcal{D}_s$ denote the available labeled source domain(s), each consisting of PSG recordings and corresponding sleep-stage annotations, and let $\mathcal{D}_t$ denote an unseen target domain with distribution shift. The objective of domain generalization is to learn a model using only $\mathcal{D}_s$ that generalizes well to such unseen domains without accessing any target-domain samples during training.

The expected risk on $\mathcal{D}_t$ can be upper bounded as~\cite{TheoryLearningDifferentBen-David10}
\begin{equation}
\epsilon_t(h) \leq \epsilon_s(h)
+ \frac{1}{2} d_{\mathcal{H}\Delta\mathcal{H}}(\mathcal{D}_s,\mathcal{D}_t)
+ \lambda,
\end{equation}
where $\epsilon_s(h)$ and $\epsilon_t(h)$ denote the source and target risks, $d_{\mathcal{H}\Delta\mathcal{H}}$ measures the domain discrepancy, and $\lambda$ represents the minimum joint risk of the shared optimal hypothesis. This bound indicates that generalization critically depends on reducing domain discrepancy and learning representations that remain invariant across domains.

Most existing domain generalization approaches reduce domain discrepancy by aligning feature distributions across multiple source domains~\cite{DeepCORALCorrelationSun16,DomainGeneralizationAdversarialLi18}. This strategy is less applicable in the practically relevant single-source setting considered here, where only one labeled dataset is available and explicit cross-domain alignment or style diversification cannot be easily performed. In addition, sleep staging introduces further challenges, as datasets vary substantially in subject populations, recording devices, and inter-scorer inconsistencies~\cite{ScoringVariabilityPolysomnographyCollop02}, making direct distribution alignment often unreliable for robust generalization.

\subsection{Spectral Signal Modeling}
Sleep stages are intrinsically characterized by distinct oscillatory rhythms that are largely shared across datasets and acquisition settings. As a result, sleep-related information tends to concentrate in a small subset of spectral components, whereas many remaining components primarily reflect nuisance or dataset-specific variations.

This structure motivates restricting the hypothesis space to focus on physiologically meaningful oscillatory components while suppressing irrelevant variations. Such a spectral-domain formulation introduces domain-consistent inductive biases that narrow $\mathcal{H}$, implicitly leading to a smaller shared optimal risk $\lambda$. By attenuating nuisance components, this restriction further reduces the cross-domain discrepancy $d_{\mathcal{H}\Delta\mathcal{H}}$.

We use Gabor filters for spectral decomposition.  Each filter with center frequency $\eta$ is defined in the time domain as
\begin{equation}\label{eq:gabor}
\psi_{\eta}(t) = a \left(
\exp\!\left(-\frac{t^2}{2\sigma^2}\right)\cos(2\pi \eta t)
- \mu
\right),
\end{equation}
where $\sigma$ controls the temporal support. The parameters $a$ and $\mu$ are chosen such that
\begin{equation}
\int \psi_{\eta}(t)\,dt = 0,\,\int \psi_{\eta}^2(t)\,dt = 1.
\end{equation}


\section{Method}
Let $\mathbf{X} \in \mathbb{R}^{N \times C \times T}$ denote PSG recordings comprising $N$ consecutive 30-second epochs, each with $C$ channels and $T$ time steps. Sleep staging aims to assign each epoch $\mathbf{X}_n$ to one of $K$ sleep stages, producing $\mathbf{Y} \in \mathbb{R}^{N \times K}$. As illustrated in Fig.~\ref{fig:framework}, our model first employs a structured spectral encoder to extract a compact embedding from each epoch, followed by a sequence encoder that captures contextual dependencies across epochs for final prediction. During training, we further adopt a mean teacher framework, in which a teacher model with exponentially averaged weights is used to enforce prediction consistency and improve generalization.

\subsection{Structured Spectral Epoch Encoder}
Without loss of generality, for a given epoch signal, we denote the time series for the $c$-th channel as $x_c(t)$. We assume that the physiological signal in each channel can be represented as a superposition of a finite number of narrowband oscillatory components, each corresponding to a canonical rhythm, and each sleep pattern as a combination of these components.

\subsubsection{Spectral Decomposition}
We implement spectral decomposition using a learnable Gabor filterbank built upon Eq.~\ref{eq:gabor}. 
Each channel is processed by an independent filterbank, and a constant-Q constraint is imposed within each channel by
\begin{equation}
\sigma(\eta)=\frac{q_c}{2\pi \eta},
\end{equation}
where $q_c$ is learnable.

Band responses are computed as
\begin{equation}
z_{c,\eta}(t)=\left|(x_c * \psi_{\eta})(t)\right|,
\label{eq:band_response}
\end{equation}
where the magnitude introduces nonlinear rectification. Each channel is equipped with an independent filterbank whose center frequencies $\{\eta_{c,j}\}_{j=1}^{F_c}$ are learnable and initialized on a logarithmic scale to provide higher resolution at lower frequencies. The responses from all channels are concatenated to form the output $\{z_i(t)\}_{i=1}^{F}$ with $F=\sum_c F_c$.

Compared with a standard 1D convolution layer with $F$ kernels of length $L$ over $C$ input channels, which requires $\mathcal{O}(FCL)$ parameters, the proposed filterbank is parameterized only by the center frequencies and one bandwidth parameter per channel, resulting in $\mathcal{O}(F + C)$ learnable parameters. The parameter count is thus independent of both the kernel size and cross-channel mixing, substantially reducing the degrees of freedom while enforcing structured spectral inductive biases.

\subsubsection{Spectral Integration}
While each bandwise signal $z_{i}(t)$ represents a narrowband component, the resulting bandwise representations may not directly reflect higher-level sleep-related structures. We therefore introduce a bandwise spectral integration mechanism to reorganize these signals into more structurally meaningful components.

We first apply band-specific temporal filtering independently to each bandwise signal $z_{i}(t)$ in order to capture characteristic temporal patterns within individual frequency bands. The filtered signals are then linearly integrated across bands to form unified temporal features:
\begin{equation}\label{eq:spectral_encoder}
\hat{z}_{m}(t) = \sum_{i=1}^{F} \alpha_{m,i}\, \bigl(w_i * z_{i}\bigr)(t),
\, m=1,\ldots,M .
\end{equation}
Here, $w_i(t)$ denotes a learnable temporal convolution kernel associated with the $i$-th frequency band, and $\{\alpha_{m,i}\}_{i=1}^F$ represents the coefficients of the $m$-th integrated component.

Since multiple integrated representations are produced, not all of them are equally informative.
We therefore introduce an adaptive calibration mechanism that automatically modulates their contributions in a sample-dependent manner.
Specifically, data-dependent scaling coefficients $\{\beta_{m}\}_{m=1}^{M}$, with $\beta_{m}\in(0,1)$, are computed from the integrated representations and used to reweight each $\hat{z}_{m}(t)$.
The calibrated representations are then aggregated to yield the output as follows:
\begin{equation}
\tilde{z}_m(t)=\beta_{m}\,\hat{z}_{m}(t),\, m=1,\ldots,M .
\end{equation}

In practice, the proposed spectral integration module is used as a drop-in replacement for standard temporal convolution layers. Combined with existing nonlinear activation functions, this design enables the construction of deep hierarchical networks while preserving the structured spectral inductive bias. 

The outputs of the epoch encoder are finally aggregated by temporal averaging to produce a fixed-dimensional representation.

\subsection{Epoch Sequence Encoder}
To capture contextual dependencies across adjacent epochs, we apply a lightweight self-attention layer~\cite{AttentionAllYouVaswani17} on top of the per-epoch representations. 
Let $\mathbf{H} = [\boldsymbol{h}_1, \dots, \boldsymbol{h}_N] \in \mathbb{R}^{N \times D}$ denote the sequence of epoch-level features, where $N$ is the number of epochs and $D$ is the feature dimension.

Positional embeddings are added to $\mathbf{H}$ to preserve epoch order, and a single-layer multi-head self-attention operation is applied to incorporate cross-epoch contextual information. 
The resulting contextualized representations are then projected through a linear classification layer with softmax activation to obtain predicted stage probabilities $\hat{\mathbf{Y}} \in \mathbb{R}^{N \times K}$.

\subsection{Spectral Consistency Regularization}
To improve robustness against spectral variability and narrowband artifacts, we introduce a spectral consistency regularization scheme that enforces prediction invariance under controlled frequency-domain perturbations~\cite{FourierbasedFrameworkDomainXu21}.

%
%
%

\subsubsection{Bandwise Spectral Perturbation}
To enhance robustness and mitigate reliance on narrowband artifacts, we introduce a bandwise spectral perturbation scheme. Given signal $x(t)$, we first compute its Fourier transform:
\begin{equation}
X(\omega) = \mathcal{F}\{x(t)\} = A(\omega) \cdot e^{j\phi(\omega)},
\end{equation}
where $A(\omega)$ and $\phi(\omega)$ denote the magnitude and phase spectra, respectively.

We then partition $A(\omega)$ into a set of predefined bands $\{ \mathcal{B}_k \}$. For each band, the magnitude is stochastically mixed with that of a randomly sampled epoch:
\begin{equation}
A_k^{\text{mix}}(\omega) = (1-\beta) A_k(\omega) +  \beta A_k'(\omega), \ \omega \in \mathcal{B}_k,
\end{equation}
where $A_k'$ is the magnitude spectrum from another epoch, and $\beta$ is the mixing coefficient. 

To further enhance temporal invariance, we add Gaussian noise to the phase spectrum within each band:
\begin{equation}
\phi_k^{\text{jitter}}(\omega) = \phi(\omega) + \epsilon(\omega), \ \epsilon(\omega) \sim \mathcal{N}(0, \sigma^2), \ \omega \in \mathcal{B}_k.
\end{equation}

The perturbed spectrum is then reconstructed and transformed back to the time domain:
\begin{equation}\label{eq:aug}
\hat{x}(t) = \mathcal{F}^{-1}\{ A^{\text{mix}}(\omega) \cdot e^{j\phi^{\text{jitter}}(\omega)} \}.
\end{equation}
To preserve signal fidelity, we set the jitter strength to $\sigma=0.02$ and sample the spectral mixing coefficient $\beta$ uniformly from the range $\in[0,0.5]$.


\subsubsection{Consistency Regularization}
To enforce prediction consistency under the proposed spectral perturbations, we adopt a momentum-based teacher–student framework~\cite{MeanTeachersAreTarvainen17}. Suppose $\mathcal{M}(\cdot;\theta)$ is the main (student) model, a momentum-based teacher $\mathcal{M}(\cdot;\theta_t)$ is updated as:
\begin{equation}
\theta_t = m \cdot \theta_t + (1 - m) \cdot \theta,
\end{equation}
where $m$ is the momentum coefficient.

Let $p_\theta(x)=\mathrm{softmax}(\mathcal{M}(x;\theta)/\tau)$ denote the predicted probability distribution. 
We minimize the symmetric KL divergence:
\begin{equation}
\mathcal{L}_{\text{cot}}
= \tau^2\left(\mathrm{KL}\left(p_\theta(\hat{x})\| p_{\theta_t}(x)\right)
+ \mathrm{KL}\left(p_\theta(x)\| p_{\theta_t}(\hat{x})\right)\right),
\end{equation}
where $\hat{x}$ denotes the perturbed signal in Eq.~\ref{eq:aug} and $\tau$ is the temperature. We adopt the default hyperparameters $\tau=10,\lambda_{\text{cot}}=2$, following the previous work~\cite{FourierbasedFrameworkDomainXu21}.

\subsection{Loss Function}
The model is trained with a combination of supervised classification and consistency regularization:
\begin{equation}
\mathcal{L}
= \mathcal{L}_{\mathrm{cls}}
+ \lambda_{\mathrm{cot}} \mathcal{L}_{\mathrm{cot}},
\label{eq:loss}
\end{equation}
where $\lambda_{\mathrm{cot}}$ balances the consistency term.

For labeled data, supervision is applied to both the original signal $x$ and its perturbed counterpart $\hat{x}$:
\begin{equation}
\mathcal{L}_{\mathrm{cls}}=\frac{1}{2}\left(
\mathrm{CE}(p_\theta(x), y)+\mathrm{CE}(p_\theta(\hat{x}), y)\right),
\end{equation}
to encourage label-consistent predictions across both views.

\section{Experiments}


\begin{table}[!t]
  \centering
  \caption{Overview of datasets used in this study.
  }
  \label{tab:dataset}
  \begin{tabular}{rcc}
  \hline
  Dataset & EEG & EOG \\
  \hline
  SleepEDF~\cite{PhysioBankPhysioToolkitPhysioNetGoldberger00}	& Fpz--Cz & Horizontal \\
  HMC~\cite{InterdatabaseValidationDeepAlvarez-Estevez21} 	& F4--M1  & E1--M2 \\
  ISRUC~\cite{ISRUCSleepComprehensivePublicKhalighi16}		& F4--M1  & E1--M2 \\
  SHHS1~\cite{SleepHeartHealthQuan97}     & C4--M1  & ROC--LOC \\
  CinC 2018~\cite{YouSnoozeYouGhassemi18}  & C3--M2  & E1--M2 \\
  \hline
  \end{tabular}
\end{table}

\begin{table*}[!t]
\centering
\caption{Single-source evaluation. Models are trained on one source domain and evaluated on all other target domains.}
\label{tab:singlesource}
\begin{tabular}{r|cccccccccc|cc}
\hline
\multirow{2}{*}{Method}
  & \multicolumn{2}{c}{SleepEDF~\cite{PhysioBankPhysioToolkitPhysioNetGoldberger00}} 
  & \multicolumn{2}{c}{HMC~\cite{InterdatabaseValidationDeepAlvarez-Estevez21}}
  & \multicolumn{2}{c}{ISRUC~\cite{ISRUCSleepComprehensivePublicKhalighi16}}
  & \multicolumn{2}{c}{SHHS1~\cite{SleepHeartHealthQuan97}}
  & \multicolumn{2}{c|}{CinC 2018~\cite{YouSnoozeYouGhassemi18}}
  & \multicolumn{2}{c}{Average} \\
 & ACC & MF1 & ACC & MF1 & ACC & MF1 & ACC & MF1 & ACC & MF1 & ACC & MF1 \\
\hline
ERM
& \underline{61.76} & 55.29 & \underline{69.56} & \underline{66.54} & 70.04 & 66.31 & 70.00 & 63.58 & 66.75 & 63.19 
& 67.62 & 62.98 \\
\hline
SAM~\cite{SharpnessawareMinimizationEfficientlyForet21} 
& 60.45 & 54.39 & 67.64 & 64.58 & \underline{71.29} & \underline{66.88} & 69.44 & 62.71 & 66.60 & 63.02 
& 67.08 & 62.32 \\
F-SAM~\cite{FriendlySharpnessAwareMinimizationLi24}
& 61.58 & \underline{55.55} & 68.91 & 65.36 & 70.27 & 65.89 & 68.81 & 62.10 & \underline{70.18} & \underline{66.04} 
& \underline{67.95} & \underline{62.99} \\
GroupDRO~\cite{DistributionallyRobustNeuralSagawa20}
& 59.24 & 55.11 & 65.35 & 64.14 & 66.76 & 64.54 & 65.95 & \underline{63.92} & 68.69 & 65.69 
& 65.20 & 62.68 \\
CIRL~\cite{CausalityInspiredRepresentationLv22}
& 51.28 & 44.64 & 63.37 & 60.02 & 66.80 & 62.62 & 69.92 & 63.15 & 63.89 & 61.19 
& 63.05 & 58.33 \\
MixStyle~\cite{DomainGeneralizationMixStyleZhou20}
& 52.59 & 46.62 & 67.83 & 64.48 & 69.57 & 65.38 & \underline{71.12} & 63.18 & 63.84 & 59.10 
& 64.99 & 59.75 \\ 
FACT~\cite{FourierbasedFrameworkDomainXu21}
& 53.97 & 48.26 & 65.40 & 61.27 & 70.11 & 65.70 & 69.20 & 60.16 & 63.54 & 59.74 
& 64.44 & 59.03 \\
\hline
SleepBand 
& \textbf{66.77} & \textbf{62.08} & \textbf{75.03} & \textbf{70.99} & \textbf{73.56} & \textbf{70.07} & \textbf{72.09} & \textbf{63.77} & \textbf{73.78} & \textbf{68.48} & \textbf{72.25} & \textbf{67.08} \\
\hline
\end{tabular}
\end{table*}

\begin{table}[!t]
\centering
\setlength{\tabcolsep}{3pt} 
\caption{Average in-domain validation performance of different methods.}
\label{tab:avg_validation}
\begin{tabular}{r|cc|cc}
\hline
& \multicolumn{2}{c|}{Source Domain} & \multicolumn{2}{c}{Target Doamin} \\
Method & ACC & F1 & ACC & F1 \\
\hline
ERM        & 83.12 $\pm$ 2.02 & 78.55 $\pm$ 3.13 & 67.62 $\pm$ 3.55 & 62.98 $\pm$ 4.56\\
SAM        & 82.84 $\pm$ 2.02 & 77.80 $\pm$ 3.41 & 67.08 $\pm$ 4.12 & 62.32 $\pm$ 4.73 \\
F-SAM      & 83.01 $\pm$ 2.01 & 78.04 $\pm$ 3.24 & 67.95 $\pm$ 3.63 & 62.99 $\pm$ 4.46 \\
Group DRO  & 81.08 $\pm$ 1.22 & 78.37 $\pm$ 2.86 & 65.20 $\pm$ 3.19 & 62.68 $\pm$ 3.82 \\
CIRL       & 83.17 $\pm$ 2.02 & 78.48 $\pm$ 3.11 & 63.05 $\pm$ 7.08 & 58.32 $\pm$ 7.75 \\
MixStyle   & 83.46 $\pm$ 1.72 & 78.84 $\pm$ 3.18 & 64.99 $\pm$ 7.44 & 59.75 $\pm$ 7.72 \\
FACT       & 82.87 $\pm$ 2.03 & 77.57 $\pm$ 4.32 & 64.44 $\pm$ 6.45 & 59.03 $\pm$ 6.47  \\
\hline
Ours       & \textbf{84.04} $\pm$ 1.44 & \textbf{79.48} $\pm$ 3.37 & \textbf{72.25} $\pm$ 3.23 & \textbf{67.08} $\pm$ 3.94 \\
\hline
\end{tabular}
\end{table}

\begin{table*}[!t]
\centering
\caption{Multi-source domain generalization. Models are trained on all but one domain and tested on the held-out target domain.}
\label{tab:lodo}
\begin{tabular}{r|cccccccccc|cc}
\hline
\multirow{2}{*}{Method}   
  & \multicolumn{2}{c}{SleepEDF~\cite{PhysioBankPhysioToolkitPhysioNetGoldberger00}} 
  & \multicolumn{2}{c}{HMC~\cite{InterdatabaseValidationDeepAlvarez-Estevez21}}
  & \multicolumn{2}{c}{ISRUC~\cite{ISRUCSleepComprehensivePublicKhalighi16}}
  & \multicolumn{2}{c}{SHHS1~\cite{SleepHeartHealthQuan97}}
  & \multicolumn{2}{c|}{CinC 2018~\cite{YouSnoozeYouGhassemi18}}
  & \multicolumn{2}{c}{Average} \\
  & ACC & MF1 & ACC & MF1 & ACC & MF1 & ACC & MF1 & ACC & MF1 & ACC & MF1 \\
\hline
ERM
& 76.67 & 71.70 & 74.03 & 72.20 & 77.61 & 73.92 & 71.41 & 64.05 & 73.91 & 69.87 & 74.72 & 70.35 \\
\hline
SAM 
& 76.95 & 71.44 & 73.86 & 71.86 & 77.61 & 74.16 & 72.92 & 65.95 & 73.73 & 69.70 & 75.01 & 70.62 \\
F-SAM
& 77.51 & 72.07 & 74.24 & 72.26 & 77.93 & 74.53 & 73.42 & 65.90 & 73.63 & 69.37 & 75.35 & 70.82 \\
MixStyle
& 77.53 & 72.39 & 72.15 & 70.83 & 78.14 & 74.74 & 70.94 & 63.54 & 72.73 & 68.49 & 74.30 & 70.00 \\
FACT 
& 78.91 & 73.34 & 74.50 & 72.12 & 77.69 & 74.71 & 73.08 & 65.01 & 73.86 & 69.41 & \underline{75.61} & 70.92 \\
CIRL 
& 76.66 & 72.14 & 73.57 & 72.03 & 78.52 & 76.16 & 69.71 & 62.51 & 72.09 & 68.53 & 74.11 & 70.27 \\
GroupDRO 
& 76.70 & 71.83 & 74.68 & 72.43 & 77.52 & 74.62 & 70.74 & 63.35 & 73.98 & 69.69 & 74.72 & 70.38 \\
\hline
MMD
& 76.17 & 71.05 & 73.75 & 71.72 & 77.62 & 74.25 & 72.19 & 64.51 & 74.20 & 69.68 & 74.79 & 70.24 \\
CORAL
& 75.88 & 70.89 & 74.27 & 72.29 & 77.70 & 74.26 & 72.84 & 65.07 & 73.74 & 69.36 & 74.89 & 70.37 \\
IRM~\cite{InvariantRiskMinimizationArjovsky20} 
& 76.78 & 71.60 & 73.60 & 71.78 & 77.99 & 74.53 & 72.01 & 64.05 & 74.51 & 71.48 & 74.98 & 70.69 \\
DANN
& 75.47 & 70.77 & 74.46 & 72.42 & 78.10 & 75.27 & 70.40 & 62.97 & 74.55 & 70.34 & 74.60 & 70.35 \\
REx
& 76.35 & 71.03 & 74.03 & 72.37 & 77.84 & 74.71 & 71.92 & 64.33 & 73.56 & 70.69 & 74.74 & 70.63 \\
\hline
SleepDG
& 77.50 & 72.35 & 74.00 & 72.18 & 78.25 & 75.03 & 73.07 & 66.04 & 73.81 & 69.80 & 75.32 & \underline{71.08} \\    
SleepBand
& 77.13 & 71.08 & 75.87 & 74.00 & 78.50 & 74.97 & 78.36 & 70.26 & 76.20 & 72.72 & \textbf{77.21} & \textbf{72.61} \\
\hline
\end{tabular}
\end{table*}

\subsection{Experimental Steup}
\textbf{Datasets.} We evaluate our method on five publicly available datasets: SleepEDF~\cite{PhysioBankPhysioToolkitPhysioNetGoldberger00}, HMC~\cite{InterdatabaseValidationDeepAlvarez-Estevez21}, ISRUC~\cite{ISRUCSleepComprehensivePublicKhalighi16}, SHHS1~\cite{SleepHeartHealthQuan97}, and CinC 2018~\cite{YouSnoozeYouGhassemi18}. 
Following~\cite{GeneralizableSleepStagingWang24}, we adopt their data processing pipeline and dataset configuration. All signals are resampled to 100 Hz, bandpass-filtered to 0.5-35 Hz, and z-score normalized. 
Each recording is segmented into sequences consisting of 20 consecutive epochs. 
The EEG and EOG channels used for each dataset are listed in Table~\ref{tab:dataset}. 


\textbf{Baselines.}
Given the limited number of sleep stage DG studies~\cite{GeneralizableSleepStagingWang24}, we primarily benchmark against established general DG baselines, categorized into single-source and multi-source settings.

For single-source settings, most existing approaches~\cite{AdvSTRevisitingDataZheng24} focus on augmenting the source domain to improve generalization. Feature-level augmentation methods, such as MixStyle~\cite{DomainGeneralizationMixStyleZhou20}, mix instance-level feature statistics (mean and variance) to simulate style variations and encourage robust feature learning. Similarly, FACT~\cite{FourierbasedFrameworkDomainXu21} augments training data by interpolating Fourier amplitude spectra between instances, enhancing robustness to distribution shifts. Other strategies optimize the training process itself to improve generalization. For instance, SAM~\cite{SharpnessawareMinimizationEfficientlyForet21} and F-SAM~\cite{FriendlySharpnessAwareMinimizationLi24} optimize for flatter loss landscapes, reducing overfitting and improving resilience to domain shifts. GroupDRO~\cite{DistributionallyRobustNeuralSagawa20} minimizes worst-group risk under distribution shifts, ensuring better performance across all data groups. Additionally, methods like CIRL~\cite{CausalityInspiredRepresentationLv22} mitigate spurious correlations by promoting invariant representations, reducing reliance on dataset-specific factors.

For multi-source settings, several approaches focus on learning domain-invariant representations. IRM~\cite{InvariantRiskMinimizationArjovsky20} enforces invariance by penalizing gradients of the loss with respect to classifier parameters across different domains, ensuring that learned features are robust to domain shifts. REx~\cite{OutofDistributionGeneralizationRiskKrueger21} minimizes the average risk while penalizing risk variance across domains, which encourages the model to focus on invariant aspects of the data that generalize well across different sources. MMD~\cite{DomainGeneralizationAdversarialLi18} aligns feature distributions across domains by using a kernel-based maximum mean discrepancy, effectively reducing domain-specific feature discrepancies. Similarly, CORAL~\cite{DeepCORALCorrelationSun16} matches second-order feature statistics (covariances) across domains, aligning the distribution of feature representations. DANN~\cite{DomainAdversarialTrainingNeuralGanin16} uses adversarial training to learn domain-invariant representations by minimizing domain classification loss, thereby encouraging the model to focus on features that are shared across domains.


\textbf{Implementation Details}:
We closely follow the architecture and optimization setup from~\cite{GeneralizableSleepStagingWang24} to ensure a fair comparison. The same temporal encoder is used. For the epoch encoder, we follow the four-block structure and dimension. Due to the lightweight design, our epoch encoder uses only approximately 23\% of the parameters of the standard version. We also use the same training setup, including the Adam optimizer with a learning rate of $10^{-3}$, weight decay of $10^{-4}$, a batch size of 32, a dropout rate of 0.1, and 50 training epochs. The input sequence length is set to 20, and the feature dimension after the epoch encoder is 512. 

\subsection{Performance on Single-source Domain Generalization}

We first evaluate single-source domain generalization by training models on one source domain and assessing their performance on all remaining unseen target domains. Quantitative results are summarized in Table~\ref{tab:singlesource}. Approaches based on loss landscape flatness, including SAM and F-SAM, yield only marginal and inconsistent improvements. In contrast, CIRL, MixStyle, and FACT consistently result in performance degradation. This trend likely reflects a fundamental mismatch between domain-agnostic generalization strategies and the low-redundancy, physiologically structured nature of sleep signals, where invariant constraints or unconstrained perturbations can suppress or distort label-defining information. Unlike generic semantic recognition tasks that benefit from redundant and style-invariant features, sleep staging relies on a small set of precise spectral and morphological patterns, making it sensitive to such operations. GroupDRO shows reduced variance and improved worst-case stability, but its conservative reweighting sacrifices average performance, indicating that robustness alone does not translate to better overall generalization in the single-domain setting. In contrast, our method consistently outperforms all baselines across every source configuration, achieving substantial improvements in average cross-domain accuracy and F1 score and demonstrating robust transferability without dependence on any specific source dataset.

To further disentangle cross-domain generalization from source-domain fitting, we report the average performance on validation splits drawn from the same source domains in Table~\ref{tab:avg_validation}. These validation splits are used solely for model selection and are disjoint from the training data. The proposed method attains validation performance comparable to that of existing baselines, indicating that its superior cross-domain results cannot be attributed to stronger optimization on the source domains.

We further report variance across datasets. In-domain variance largely reflects intrinsic dataset difficulty, whereas additional variance under cross-domain evaluation may be associated with sensitivity to spurious, dataset-dependent correlations. Comparing Table~\ref{tab:singlesource} and Table~\ref{tab:avg_validation}, our method shows relatively low additional variance under cross-domain evaluation, further indicating limited domain-specific bias.

These observations point to the distinctive nature of sleep stage classification as a domain generalization problem. Rather than benefiting from generic invariance-inducing strategies, effective generalization appears to hinge on preserving task-specific physiological structures that govern the underlying sleep dynamics.

\subsection{Performance on Multi-source Domain Generalization}
Although our method is not explicitly designed for multi-source domain generalization, we further evaluate its effectiveness under a multi-source setting to assess its robustness to cross-domain distribution shifts. We adopt a leave-one-domain-out evaluation protocol, where models are trained on multiple source domains and evaluated on a held-out target domain. Training on diverse source domains enables the model to perform additional cross-domain comparisons, which helps reduce reliance on domain-specific patterns and encourages the learning of more generalizable representations.

Table~\ref{tab:lodo} reports the results on five public sleep datasets. The base model denotes the shared backbone trained without any domain generalization strategy and serves as a reference baseline. Compared to this baseline, our method consistently improves performance across all held-out domains. In particular, it achieves the highest average accuracy and macro-F1 score among all compared methods, demonstrating strong generalization capability under domain shifts.

We further observe that several generic DG approaches yield only marginal improvements over the base model, and in some cases even degrade performance (e.g., DANN). This suggests that conventional domain generalization techniques may be suboptimal for sleep staging, where domain discrepancies often arise from complex spectral and recording-related variations rather than explicit domain semantics.

\begin{table}[!t]
\centering
\caption{Ablation study of different components in the proposed method.}
\label{tab:ablation}
\setlength{\tabcolsep}{5pt}
\begin{tabular}{c|ccc|cc|cc}
\hline
& Gabor & Spectral & Mean & \multicolumn{2}{c|}{Single-source} & \multicolumn{2}{c}{Multi-source}  \\ 
& Filter & Integration & Teacher & ACC & MF1 & ACC & MF1\\
\hline
(a) & -- & -- & --  & 67.62 & 62.98 & 74.72 & 70.35  \\
(b) & -- & $\checkmark$ & -- & 60.31 & 55.58 & 68.42 & 63.81 \\
(c) & -- & -- & $\checkmark$ & 67.82 & 61.99 & 76.08 & 70.99 \\
(d) & -- & $\checkmark$ & $\checkmark$ & 58.74 & 52.45 & 71.32 & 65.15 \\
\hline
(e) & $\checkmark$ & -- & -- & 69.14 & 64.34 & 76.09 & 71.60 \\
(f) & $\checkmark$ & $\checkmark$ & -- & 69.14 & 64.16 & 75.40 & 71.05 \\
(g) & $\checkmark$ & -- & $\checkmark$ & 71.55 & 66.19 & 76.27 & 71.85 \\
(h) & $\checkmark$ & $\checkmark$ & $\checkmark$ & \textbf{71.71} & \textbf{67.33} & \textbf{77.21} & \textbf{72.61} \\
\hline
\end{tabular}
\end{table}

\begin{table}[!t]
\centering
\caption{Performance with different bandpass filters.}
\label{tab:sinc}
\begin{tabular}{r|cc|cc}
\hline
\multirow{2}{*}{Method} & \multicolumn{2}{c|}{Single-source} & \multicolumn{2}{c}{Multi-source}  \\ 
& ACC & MF1 & ACC & MF1\\
\hline
Sinc Filter  & 65.27 & 59.76 & 76.21 & 71.79 \\
Gabor Filter & 71.71 & 67.33 & 77.21 & 72.61 \\
\hline
Improvement  & 6.44 & 7.57 & 0.99 & 0.82 \\
\hline
\end{tabular}
\end{table}

\subsection{Ablation Study}
In this section, we conduct ablation studies to assess the contribution of each module in our framework. Results are reported in Table~\ref{tab:ablation}. The best performance is achieved only when all three components are jointly enabled. Individual modules or partial combinations provide limited improvements, while their integration yields consistent and significant gains, demonstrating strong complementarity and synergy among the proposed designs.

\subsubsection{Gabor Filter}
Incorporating the Gabor filter consistently improves performance across all settings. When used alone, it already surpasses the baseline in both multi-source and single-source scenarios. Moreover, all configurations that include the Gabor module achieve better results than their counterparts without it, demonstrating that Gabor filtering provides stable and discriminative spatial representations and serves as a robust performance booster for the overall framework.

Moreover, we replace the Gabor filter with a learnable bandpass filter implemented using parameterized Sinc functions, which forms a differentiable finite impulse response (FIR) filter whose cutoff frequencies are adapted during training~\cite{SpeakerRecognitionRawRavanelli19}. The performance is shown in Table~\ref{tab:sinc}. In multi-source settings, while slightly inferior to the Gabor filter, using the Sinc filter still outperforms SleepDG, indicating that bandpass filtering provides useful inductive biases. However, in single-source settings, performance degrades, suggesting that bandpass filtering alone offers insufficient regularization when only a single data source is available.

\subsubsection{Spectral Integration}
The proposed spectral integration adopts a parameter-efficient bandwise filtering strategy that imposes strong spectral inductive bias. However, due to its constrained formulation and limited cross-band interactions, the module alone is insufficient to learn highly discriminative representations, often resulting in degraded performance. Instead, it relies on other spatial or temporal modeling components to provide effective structural guidance. When jointly optimized with these modules, spectral integration supplies complementary frequency-domain cues that enhance the learned representations, leading to consistent improvements and the best performance in the full configuration.

\subsubsection{Mean Teacher}
The mean teacher brings modest improvements in the multi-source setting but provides limited gains for single-source training. This is expected since consistency regularization relies on sufficient data diversity and meaningful feature perturbations. When combined with the Gabor and spectral modules, which produce more structured representations, the consistency constraint becomes more effective, leading to more noticeable performance improvements. This indicates that the mean teacher mainly serves as a complementary regularizer rather than a standalone contributor.

To analyze the performance of our bandwise perturbation, we compare the performance of model variant c in Table~\ref{tab:ablation} with FACT, where the primary difference lies in the augmentation method. The results show that in the multi-source setting, our method performs similarly to FACT, with only limited improvement. Under the single-source setting, our augmentation strategy, which constrains excessive variability through physiologically informed perturbations, leads to a marked performance gain. It achieves an improvement of 2.96 in F1 and 3.38 in accuracy. However, the mean teacher alone still performs similarly to the baseline. This indicates that although we avoid the performance drop, its inherent consistency constraint is insufficient and also underscores that our design of these components constitutes a collaborative strategy.

\begin{figure*}[!t]
    \centering
    \subfloat[]{\includegraphics[width=0.33\linewidth]{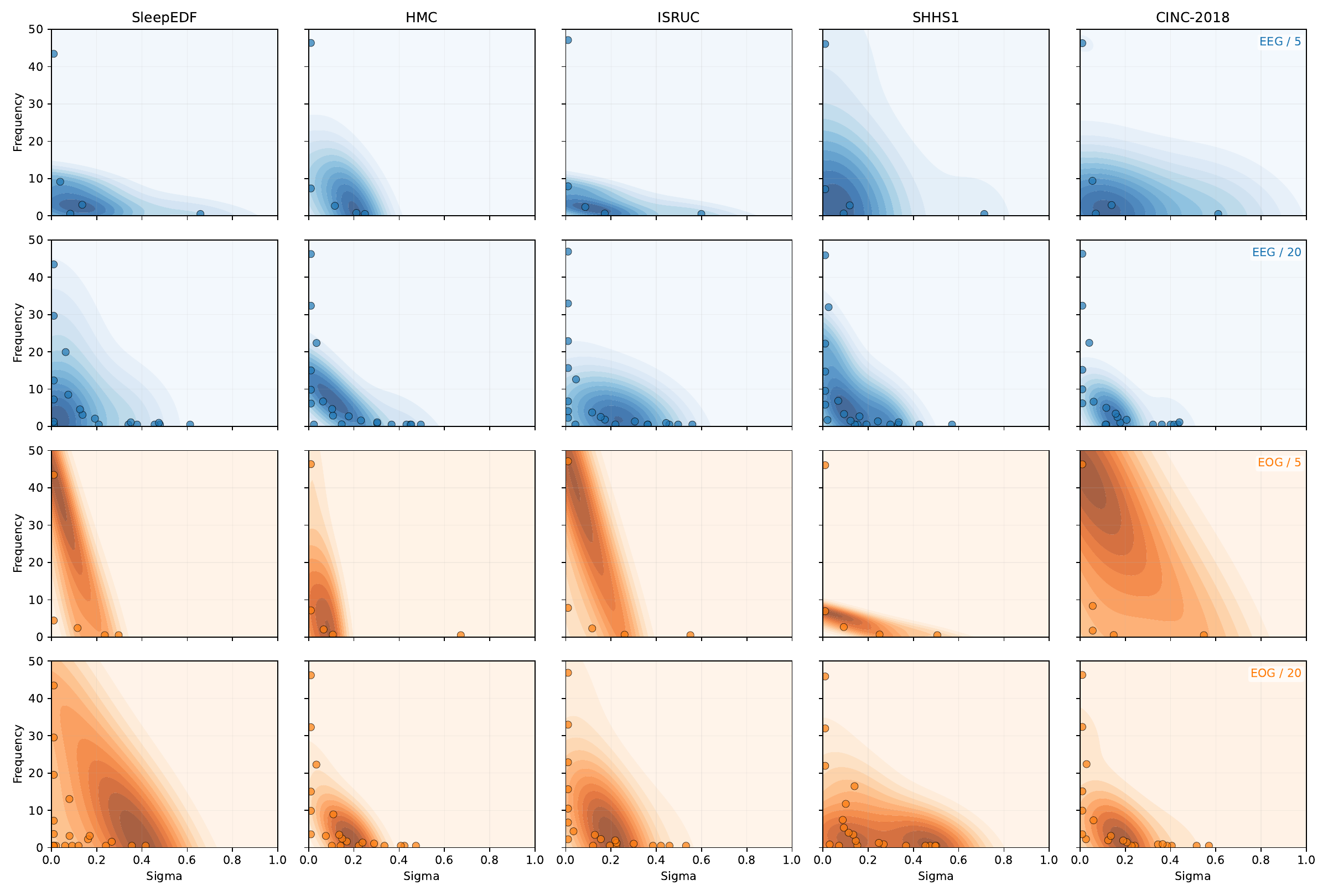}%
        \label{fig:learned_filter_patterns_in_domain}}%
    \hfil
    \subfloat[]{\includegraphics[width=0.33\linewidth]{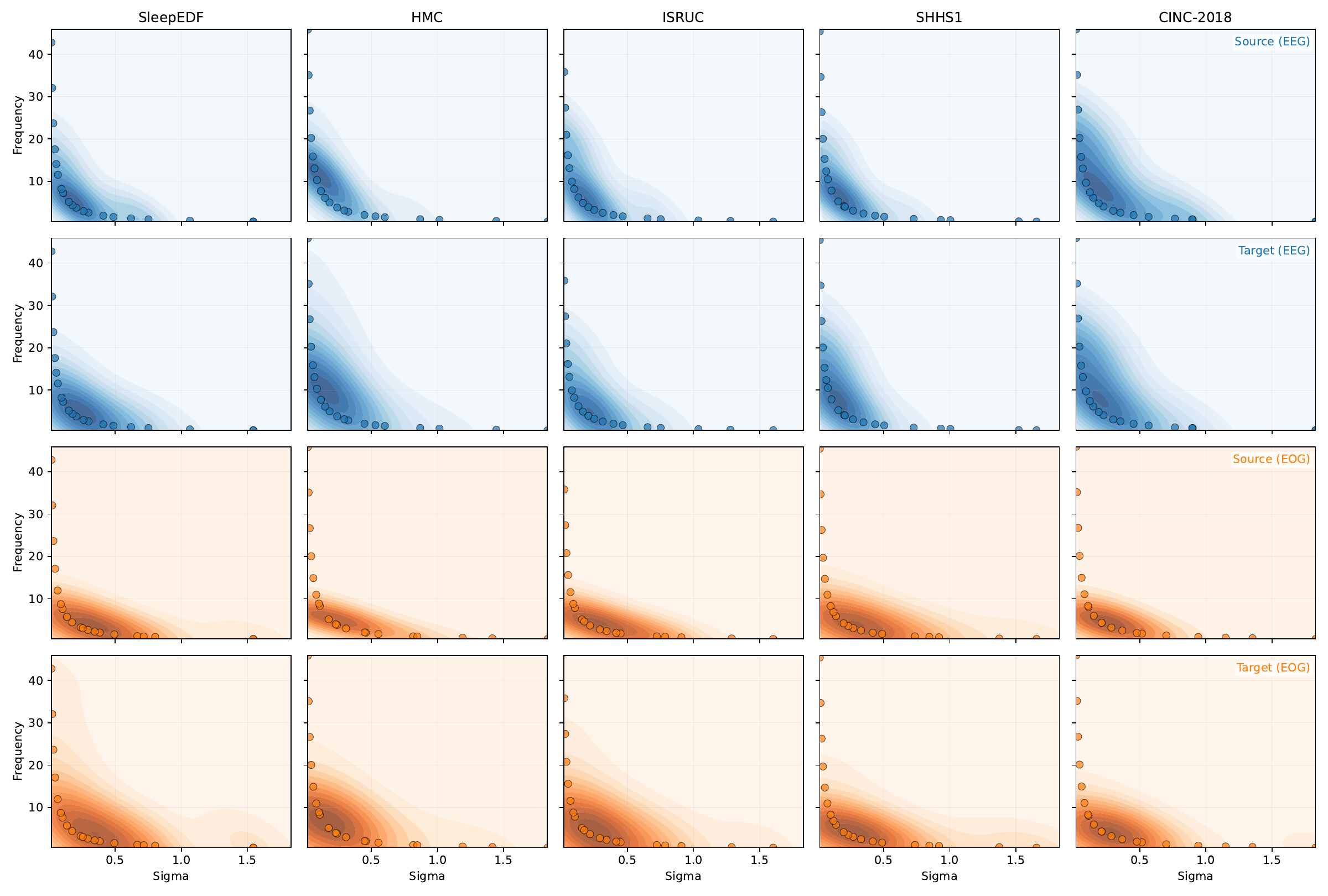}%
        \label{fig:learned_filter_patterns_single_source}}%
    \hfil
    \subfloat[]{\includegraphics[width=0.33\linewidth]{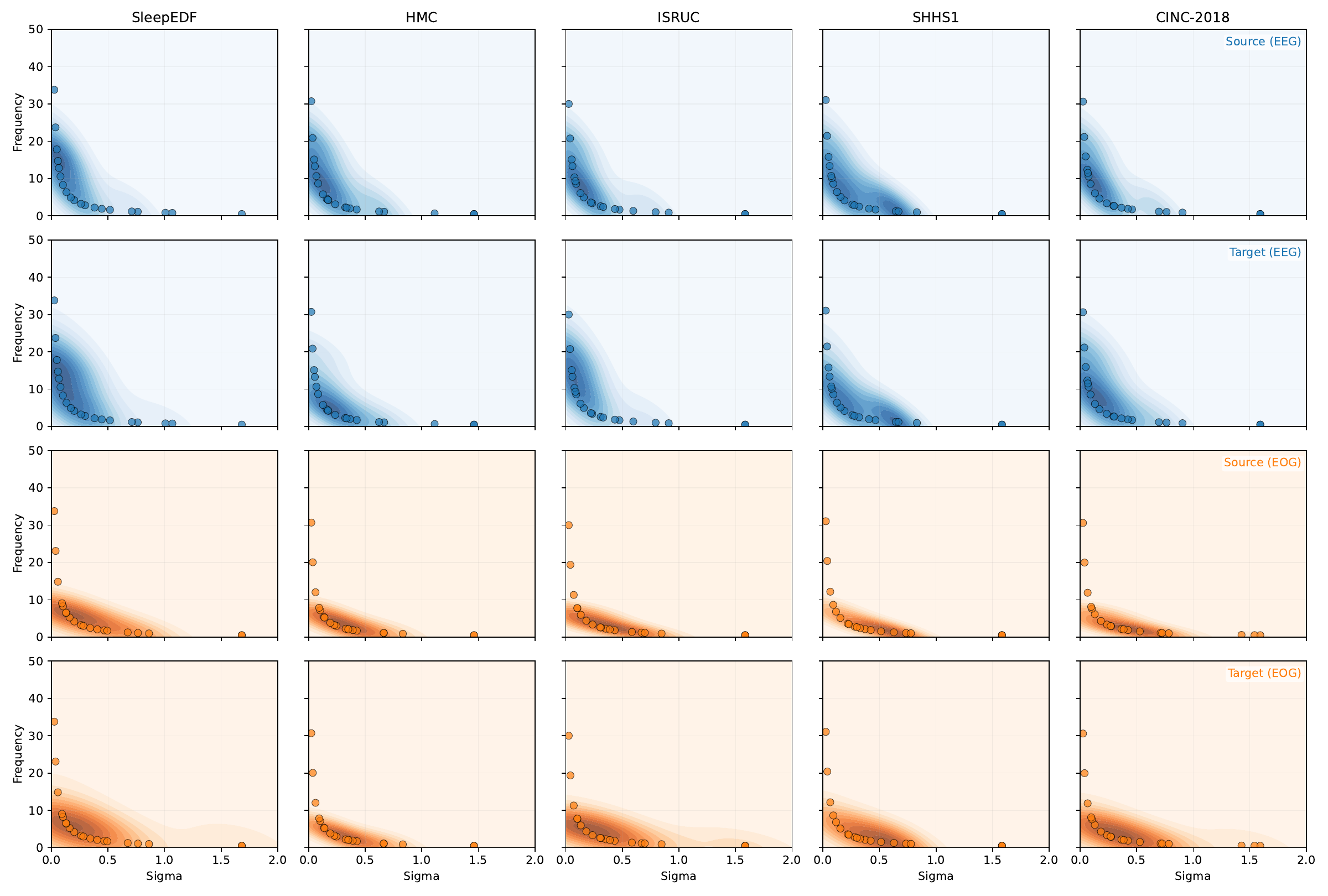}%
        \label{fig:learned_filter_patterns_multi_source}}%
    \caption{Learned filters for different signal channels (e.g., EEG and EOG) across datasets with importance-weighted kernel density estimates. (a) Learned Gabor filters from a single source domain, with importance computed based on the target domain's performance. (b) Learned constant-Q Gabor filters from the same source domain, with importance computed using both the source and target domain performances.}
    \label{fig:learned_filter_patterns}
\end{figure*}

\begin{figure}[!t]
    \centering
    \includegraphics[width=0.98\linewidth]{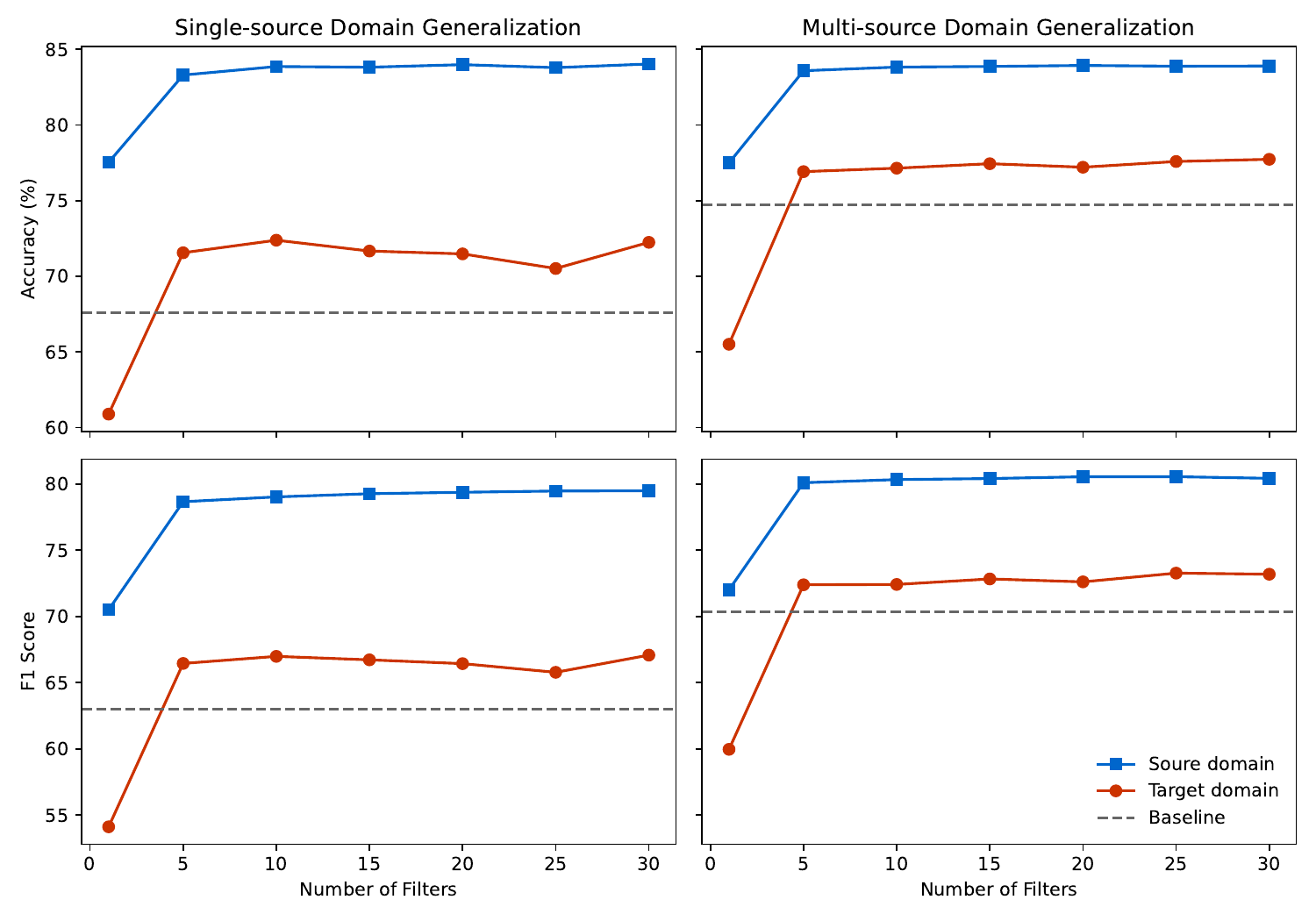}
    \caption{Performance on different numbers of filters.}
   \label{fig:number_of_filters}
\end{figure}

\begin{table}[!t]
\centering
\caption{Performance with different bandpass filters.}
\label{tab:standard_garbor}
\begin{tabular}{r|cc|cc}
\hline
\multirow{2}{*}{Method} & \multicolumn{2}{c|}{Single-source} & \multicolumn{2}{c}{Multi-source}  \\ 
& ACC & MF1 & ACC & MF1\\
\hline
w/o Constant-Q & 69.16 & 64.16 & 77.10 & 72.53 \\
w Constant-Q & 71.71 & 67.33 & 77.21 & 72.61 \\
\hline
\end{tabular}
\end{table}

\subsection{Visualization of Learned Filters}
To facilitate the interpretation of the learned representations, we visualize the learned Gabor filters in Fig.~\ref{fig:learned_filter_patterns}, where each marker corresponds to a filter. The importance of each filter is determined via an ablation study, where we assess the performance drop resulting from the removal of each filter. This performance drop serves as a measure of the filter's importance. The filters are then visualized by plotting the weighted kernel density based on their computed importance scores.

We first train models on individual datasets to evaluate in-domain performance, highlighting dataset-specific differences. Fig.~\ref{fig:learned_filter_patterns_in_domain} shows models trained with standard Gabor filters (e.g., 5 or 20 filters). Analysis reveals significant heterogeneity in the learned filter parameters across datasets, reflecting the model's adaptive focus on distinct oscillatory features. Despite all datasets containing waveforms defined in clinical guidelines for sleep staging, cross-dataset differences persist. However, with a sufficient number of filters, the learned filters converge in parameter space, clustering in regions with low center frequency and narrow bandwidth. While filter importance varies, the spatial distribution becomes less dispersed with more filters, suggesting a core subset of transferable filters. Motivated by this finding, we employ constant-q Gabor filters to maximize such cross-dataset transferability.

As shown in Fig.~\ref{fig:learned_filter_patterns_single_source} and Fig.~\ref{fig:learned_filter_patterns_multi_source}, the learned filters exhibit high consistency across different datasets when a constant-Q constraint is imposed. Furthermore, the importance of these filters also demonstrates strong agreement between the source and target domains. In the multi-source setting, the increased data variability leads to a further regularization of the learned filter patterns compared to the single-source scenario as expected.

\subsection{Number of Gabor Filters} 
We study the effect of the number of Gabor filters on model performance, as shown in Fig.~\ref{fig:number_of_filters}. Using only one filter per channel leads to degraded results due to insufficient time-frequency coverage, which fails to capture the diverse oscillatory patterns in sleep signals. When a sufficient number of filters is used, the performance quickly surpasses the baseline. We choose the configuration with the highest in-domain accuracy to report the general performance. Specifically, for single-source domain generalization, we select 30 filters per channel, and for multi-source domain generalization, we select 20 filters per channel. However, it can also be observed that our method is not highly sensitive to the exact number of filters once a moderate number of filters is used. Therefore, for simplicity, we use 20 filters as the default in other experiments.

\begin{table}[!t]
\centering
\caption{Computation and parameter efficiency comparison between the baseline and the proposed SleepBand. FLOPs and parameter counts are reported for both the epoch encoder and the full model.}
\label{tab:efficiency_comparison}
\begin{tabular}{r|cc|cc}
\hline
\multirow{2}{*}{Method} & \multicolumn{2}{c|}{Epoch Encoder} & \multicolumn{2}{c}{Full Model} \\ 
& FLOPs (G) & Params (M) & FLOPs (G) & Params (M)\\
\hline
Baseline & 2.67 &  1.56 & 2.70 & 2.35 \\
SleepBand &  0.36 & 0.36 & 0.63 & 1.15 \\
\hline
Reduction & 86.5\% & 76.9\% & {76.7\%} & {51.1\%} \\
\hline
\end{tabular}
\end{table}
\subsection{Computation and parameter efficiency}
To further demonstrate the efficiency of our approach, we compare the computational cost and parameter count of SleepBand against a conventional convolutional baseline. As summarized in Table~\ref{tab:efficiency_comparison}, SleepBand drastically reduces both FLOPs and parameters at the epoch encoder level by 86.5\% and 76.9\%, respectively. For the full model, the reductions remain substantial, reaching 76.7\% in FLOPs and 51.1\% in parameters. These gains highlight how the structured, interpretable Gabor filter bank not only enhances cross‑dataset consistency but also yields a highly lightweight architecture, making it suitable for deployment in resource‑constrained and real‑time monitoring environments.

\section{Conclusion}
Domain generalization for sleep staging presents unique challenges due to the non-stationary nature and reliance on subtle, physiology-specific inference. Existing DG methods often underperform, as they are not tailored to the structured, rule-based nature of sleep scoring and may overfit to domain-specific artifacts. This work introduces a physiologically informed framework that guides the model to focus on sleep-relevant oscillatory patterns through spectral inductive biases. Our method achieves robust performance across domains, including in single-domain settings, demonstrating the value of domain-agnostic design grounded in signal characteristics. Code will be released.

\bibliographystyle{IEEEtran}
\bibliography{references.bib}

@article{DomainAdversarialTrainingNeuralGanin16,
  title = {Domain-Adversarial Training of Neural Networks},
  author = {Ganin, Yaroslav and Ustinova, Evgeniya and Ajakan, Hana and Germain, Pascal and Larochelle, Hugo and Laviolette, Fran{\c c}ois and Marchand, Mario and Lempitsky, Victor},
  year = 2016,
  journal = {Journal of Machine Learning Research},
  volume = {17},
  number = {1},
  pages = {2096--2030},
  doi = {10.48550/arXiv.1505.07818}
}

@inproceedings{MeanTeachersAreTarvainen17,
  title = {Mean Teachers Are Better Role Models: Weight-Averaged Consistency Targets Improve Semi-Supervised Deep Learning Results},
  booktitle = {Annual {{Conference}} on {{Neural Information Processing Systems}}},
  author = {Tarvainen, Antti and Valpola, Harri},
  year = 2017,
  volume = {30},
  pages = {1195--1204}
}

@inproceedings{DeepCORALCorrelationSun16,
  title = {Deep {{CORAL}}: Correlation Alignment for Deep Domain Adaptation},
  booktitle = {{{ECCV}}},
  author = {Sun, Baochen and Saenko, Kate},
  editor = {Hua, Gang and J{\'e}gou, Herv{\'e}},
  year = 2016,
  volume = {9915},
  pages = {443--450}
}

@inproceedings{AttentionAllYouVaswani17,
  title = {Attention Is All You Need},
  booktitle = {Annual {{Conference}} on {{Neural Information Processing Systems}}},
  author = {Vaswani, Ashish and Shazeer, Noam M. and Parmar, Niki and Uszkoreit, Jakob and Jones, Llion and Gomez, Aidan N. and Kaiser, Lukasz and Polosukhin, Illia},
  year = 2017,
  volume = {30},
  pages = {5998--6008}
}

@inproceedings{YouSnoozeYouGhassemi18,
  title = {You {{Snooze}}, {{You Win}}: The {{PhysioNet}}/{{Computing}} in {{Cardiology Challenge}} 2018},
  booktitle = {2018 {{Computing}} in {{Cardiology Conference}} ({{CinC}})},
  author = {Ghassemi, Mohammad M and Moody, Benjamin E and Lehman, Li-Wei H and Song, Christopher and Li, Qiao and Sun, Haoqi and Mark, Roger G and Westover, M Brandon and Clifford, Gari D},
  year = 2018,
  volume = {45},
  pages = {1--4},
  doi = {10.22489/CinC.2018.049}
}

@article{TransferableSelfSupervisedInstanceZhao22,
  title = {Transferable {{Self-Supervised Instance Learning}} for {{Sleep Recognition}}},
  author = {Zhao, Aite and Wang, Yue and Li, Jianbo},
  year = 2022,
  journal = {IEEE Transactions on Multimedia},
  volume = {25},
  pages = {4464--4477},
  doi = {10.1109/TMM.2022.3176751}
}

@article{TheoryLearningDifferentBen-David10,
  title = {A Theory of Learning from Different Domains},
  author = {{Ben-David}, Shai and Blitzer, John and Crammer, K. and Kulesza, Alex and Pereira, Fernando C and Vaughan, Jennifer Wortman},
  year = 2010,
  journal = {Machine Learning},
  volume = {79},
  number = {1-2},
  pages = {151--175},
  doi = {10.1007/s10994-009-5152-4}
}

@article{PhysioBankPhysioToolkitPhysioNetGoldberger00,
  title = {{{PhysioBank}}, {{PhysioToolkit}}, and {{PhysioNet}}: {{Components}} of a {{New Research Resource}} for {{Complex Physiologic Signals}}},
  author = {Goldberger, Ary L. and Amaral, Luis A. N. and Glass, Leon and Hausdorff, Jeffrey M. and Ivanov, Plamen Ch. and Mark, Roger G. and Mietus, Joseph E. and Moody, George B. and Peng, Chung-Kang and Stanley, H. Eugene},
  year = 2000,
  journal = {Circulation},
  volume = {101},
  number = {23},
  pages = {e215-e220},
  doi = {10.1161/01.CIR.101.23.e215}
}

@article{DeepSleepNetModelAutomaticSupratak17,
  title = {{{DeepSleepNet}}: {{A Model}} for {{Automatic Sleep Stage Scoring Based}} on {{Raw Single-Channel EEG}}},
  author = {Supratak, Akara and Dong, Hao and Wu, Chao and Guo, Yike},
  year = 2017,
  journal = {IEEE Transactions on Neural Systems and Rehabilitation Engineering},
  volume = {25},
  number = {11},
  pages = {1998--2008},
  doi = {10.1109/TNSRE.2017.2721116}
}

@article{SleepTransformerAutomaticSleepPhan22,
  title = {{{SleepTransformer}}: {{Automatic Sleep Staging With Interpretability}} and {{Uncertainty Quantification}}},
  author = {Phan, Huy and Mikkelsen, Kaare and Chen, Oliver Y. and Koch, Philipp and Mertins, Alfred and De Vos, Maarten},
  year = 2022,
  journal = {IEEE Transactions on Biomedical Engineering},
  volume = {69},
  number = {8},
  pages = {2456--2467},
  doi = {10.1109/TBME.2022.3147187}
}

@article{PersonalizationAutomaticSleepLorenzen24,
  title = {Personalization of Automatic Sleep Scoring: How Best to Adapt Models to Personal Domains in Wearable {{EEG}}},
  author = {Lorenzen, Kristian P. and Heremans, Elisabeth R. M. and {de Vos}, Maarten and Mikkelsen, Kaare B.},
  year = 2024,
  journal = {IEEE Journal of Biomedical and Health Informatics},
  volume = {28},
  number = {10},
  pages = {5804--5815},
  doi = {10.1109/JBHI.2024.3409165}
}

@inproceedings{LearningUnifiedModelYe24,
  title = {Learning Unified Model for Sleep Health Monitoring},
  booktitle = {2024 8th {{International Conference}} on {{Communication}} and {{Information Systems}} ({{ICCIS}})},
  author = {Ye, Gaohan and Lu, Zhi and Zhang, Dongheng and Zhou, Fang and Song, Ruiyuan and Shu, Lingjie and Pu, Yu and Chen, Yan},
  year = 2024,
  pages = {27--32},
  doi = {10.1109/ICCIS63642.2024.10779418}
}

@inproceedings{FourierbasedFrameworkDomainXu21,
  title = {A {{Fourier-based Framework}} for {{Domain Generalization}}},
  booktitle = {2021 {{IEEE}}/{{CVF Conference}} on {{Computer Vision}} and {{Pattern Recognition}} ({{CVPR}})},
  author = {Xu, Qinwei and Zhang, Ruipeng and Zhang, Ya and Wang, Yanfeng and Tian, Qi},
  year = 2021,
  pages = {14378--14387},
  doi = {10.1109/CVPR46437.2021.01415}
}

@article{SleepHeartHealthQuan97,
  title = {The Sleep Heart Health Study: Design, Rationale, and Methods},
  author = {Quan, Stuart F. and Howard, Barbara V. and Iber, Conrad and Kiley, James P. and Nieto, F. Javier and O'Connor, George T. and Rapoport, David M. and Redline, Susan and Robbins, John and Samet, Jonathan M. and Wahl, {\ddag}Patricia W.},
  year = 1997,
  journal = {Sleep},
  volume = {20},
  number = {12},
  pages = {1077--1085},
  doi = {10.1093/sleep/20.12.1077}
}

@inproceedings{GeneralizableSleepStagingWang24,
  title = {Generalizable Sleep Staging via Multi-Level Domain Alignment},
  booktitle = {Proceedings of the {{AAAI Conference}} on {{Artificial Intelligence}}},
  author = {Wang, Jiquan and Zhao, Sha and Jiang, Haiteng and Li, Shijian and Li, Tao and Pan, Gang},
  year = 2024,
  eprint = {2401.05363},
  primaryclass = {eess},
  pages = {265--273},
  doi = {10.1609/aaai.v38i1.27779}
}

@article{ISRUCSleepComprehensivePublicKhalighi16,
  title = {{{ISRUC-sleep}}: A Comprehensive Public Dataset for Sleep Researchers},
  author = {Khalighi, Sirvan and Sousa, Teresa and Santos, Jos{\'e} Moutinho and Nunes, Urbano},
  year = 2016,
  journal = {Computer Methods and Programs in Biomedicine},
  volume = {124},
  pages = {180--192},
  doi = {10.1016/j.cmpb.2015.10.013}
}

@inproceedings{ATTAAdaptiveTestTimeJia24,
  title = {{{ATTA}}: Adaptive Test-Time Adaptation for Multi-Modal Sleep Stage Classification},
  booktitle = {Proceedings of the {{Thirty-Third International Joint Conference}} on {{Artificial Intelligence}}},
  author = {Jia, Ziyu and Yang, Xihao and Zhou, Chenyang and Deng, Haoyang and Jiang, Tianzi},
  year = 2024,
  pages = {5882--5890},
  doi = {10.24963/ijcai.2024/650}
}

@article{SleepSpindlesElectrographicCoppieterstWallant16,
  title = {Sleep Spindles as an Electrographic Element: Description and Automatic Detection Methods},
  author = {Coppieters 't Wallant, Doroth{\'e}e and Maquet, Pierre and Phillips, Christophe},
  year = 2016,
  journal = {Neural Plasticity},
  volume = {2016},
  pages = {6783812},
  doi = {10.1155/2016/6783812},
  pmcid = {PMC4958487},
  pmid = {27478649}
}

@article{DomainGeneralizationSurveyZhou23,
  title = {Domain Generalization: A Survey},
  author = {Zhou, Kaiyang and Liu, Ziwei and Qiao, Yu and Xiang, Tao and Loy, Chen Change},
  year = 2023,
  journal = {IEEE Transactions on Pattern Analysis and Machine Intelligence},
  volume = {45},
  number = {4},
  pages = {4396--4415},
  doi = {10.1109/TPAMI.2022.3195549}
}

@inproceedings{SharpnessawareMinimizationEfficientlyForet21,
  title = {Sharpness-Aware Minimization for Efficiently Improving Generalization},
  booktitle = {Proceedings of the 9th {{International Conference}} on {{Learning Representations}}},
  author = {Foret, Pierre and Kleiner, Ariel and Mobahi, Hossein and Neyshabur, Behnam},
  year = 2021,
  eprint = {2010.01412},
  primaryclass = {cs},
  doi = {10.48550/arXiv.2010.01412}
}

@inproceedings{DomainGeneralizationAdversarialLi18,
  title = {Domain Generalization with Adversarial Feature Learning},
  booktitle = {2018 {{IEEE}}/{{CVF Conference}} on {{Computer Vision}} and {{Pattern Recognition}}},
  author = {Li, Haoliang and Pan, Sinno Jialin and Wang, Shiqi and Kot, Alex C.},
  year = 2018,
  pages = {5400--5409},
  doi = {10.1109/CVPR.2018.00566}
}

@inproceedings{SpeakerRecognitionRawRavanelli19,
  title = {Speaker {{Recognition}} from {{Raw Waveform}} with {{SincNet}}},
  booktitle = {2018 {{IEEE Spoken Language Technology Workshop}} ({{SLT}})},
  author = {Ravanelli, Mirco and Bengio, Yoshua},
  year = 2018,
  eprint = {1808.00158},
  primaryclass = {eess},
  doi = {10.1109/slt.2018.8639585}
}

@inproceedings{SIMPLESpecializedModelSampleLi23,
  title = {{{SIMPLE}}: Specialized Model-Sample Matching for Domain Generalization},
  booktitle = {The {{Eleventh International Conference}} on {{Learning Representations}}},
  author = {Li, Ziyue and Ren, Kan and Jiang, Xinyang and Shen, Yifei and Zhang, Haipeng and Li, Dongsheng},
  year = 2023
}

@inproceedings{PriorSleepNetEnhancingSleepShu25,
  title = {{{PriorSleepNet}}: Enhancing Sleep Stage Classification through Prior Knowledge Integration},
  booktitle = {5th {{International Conference}} on {{Consumer Electronics}} and {{Computer Engineering}}},
  author = {Shu, Lingjie and Zhang, Dongheng and Ye, Gaohan and Wu, Zhi and Lu, Zhi and Zhou, Fang and Pu, Yu and Chen, Yan},
  year = 2025,
  pages = {476--481},
  doi = {10.1109/ICCECE65250.2025.10984605}
}

@article{InterdatabaseValidationDeepAlvarez-Estevez21,
  title = {Inter-Database Validation of a Deep Learning Approach for Automatic Sleep Scoring},
  author = {{Alvarez-Estevez}, Diego and Rijsman, Roselyne M.},
  year = 2021,
  journal = {PLOS One},
  volume = {16},
  number = {8},
  pages = {e0256111},
  doi = {10.1371/journal.pone.0256111}
}

@inproceedings{LearningLearnSingleQiao20,
  title = {Learning to Learn Single Domain Generalization},
  booktitle = {2020 {{IEEE}}/{{CVF Conference}} on {{Computer Vision}} and {{Pattern Recognition}} ({{CVPR}})},
  author = {Qiao, Fengchun and Zhao, Long and Peng, Xi},
  year = 2020,
  pages = {12553--12562},
  doi = {10.1109/CVPR42600.2020.01257}
}

@inproceedings{DomainGeneralizationConditionalLiu21,
  title = {Domain Generalization under Conditional and Label Shifts via Variational Bayesian Inference},
  booktitle = {Proceedings of the {{Thirtieth International Joint Conference}} on {{Artificial Intelligence}}},
  author = {Liu, Xiaofeng and Hu, Bo and Jin, Linghao and Han, Xu and Xing, Fangxu and Ouyang, Jinsong and Lu, Jun and El Fakhri, Georges and Woo, Jonghye},
  year = 2021,
  eprint = {2107.10931},
  primaryclass = {cs},
  pages = {881--887},
  doi = {10.24963/ijcai.2021/122}
}

@misc{InvariantRiskMinimizationArjovsky20,
  title = {Invariant Risk Minimization},
  author = {Arjovsky, Martin and Bottou, L{\'e}on and Gulrajani, Ishaan and {Lopez-Paz}, David},
  year = 2020,
  number = {arXiv:1907.02893},
  eprint = {1907.02893},
  primaryclass = {stat},
  doi = {10.48550/arXiv.1907.02893}
}

@inproceedings{CausalityInspiredRepresentationLv22,
  title = {Causality Inspired Representation Learning for Domain Generalization},
  booktitle = {{{IEEE}}/{{CVF Conference}} on {{Computer Vision}} and {{Pattern Recognition}} ({{CVPR}})},
  author = {Lv, Fangrui and Liang, Jian and Li, Shuang and Zang, Bin and Liu, Chi Harold and Wang, Ziteng and Liu, Di},
  year = 2022,
  pages = {8036--8046},
  doi = {10.1109/CVPR52688.2022.00788}
}

@inproceedings{DistributionallyRobustNeuralSagawa20,
  title = {Distributionally Robust Neural Networks for Group Shifts: On the Importance of Regularization for Worst-Case Generalization},
  booktitle = {{{ICLR}}},
  author = {Sagawa, Shiori and Koh, Pang Wei and Hashimoto, Tatsunori B. and Liang, Percy},
  year = 2020,
  eprint = {1911.08731},
  primaryclass = {cs},
  doi = {10.48550/arXiv.1911.08731}
}

@inproceedings{DomainGeneralizationMixStyleZhou20,
  title = {Domain Generalization with {{MixStyle}}},
  booktitle = {International {{Conference}} on {{Learning Representations}}},
  author = {Zhou, Kaiyang and Yang, Yongxin and Qiao, Yu and Xiang, Tao},
  year = 2020
}

@inproceedings{OutofDistributionGeneralizationRiskKrueger21,
  title = {Out-of-Distribution Generalization via Risk Extrapolation},
  booktitle = {International {{Conference}} on {{Machine Learning}}},
  author = {Krueger, David and Caballero, Ethan and Jacobsen, J{\"o}rn-Henrik and Zhang, Amy and Binas, Jonathan and Zhang, Dinghuai and Priol, R{\'e}mi Le and Courville, Aaron C.},
  editor = {Meila, Marina and Zhang, Tong},
  year = 2021,
  series = {Proceedings of {{Machine Learning Research}}},
  volume = {139},
  eprint = {2003.00688},
  primaryclass = {cs},
  pages = {5815--5826}
}

@article{RobustSleepNetTransferLearningGuillot21,
  title = {{{RobustSleepNet}}: {{Transfer}} Learning for Automated Sleep Staging at Scale},
  author = {Guillot, Antoine and Thorey, Valentin},
  year = 2021,
  journal = {IEEE Transactions on Neural Systems and Rehabilitation Engineering},
  volume = {29},
  eprint = {2101.02452},
  primaryclass = {stat},
  pages = {1441--1451},
  doi = {10.1109/TNSRE.2021.3098968}
}

@article{ADASTAttentiveCrossdomainEldele23,
  title = {{{ADAST}}: {{Attentive}} Cross-Domain {{EEG-based}} Sleep Staging Framework with Iterative Self-Training},
  author = {Eldele, Emadeldeen and Ragab, Mohamed and Chen, Zhenghua and Wu, Min and Kwoh, Chee-Keong and Li, Xiaoli and Guan, Cuntai},
  year = 2023,
  journal = {IEEE Transactions on Emerging Topics in Computational Intelligence},
  volume = {7},
  number = {1},
  eprint = {2107.04470},
  primaryclass = {cs},
  pages = {210--221},
  doi = {10.1109/TETCI.2022.3189695}
}

@article{DeepTransferLearningRadha21,
  title = {A Deep Transfer Learning Approach for Wearable Sleep Stage Classification with Photoplethysmography},
  author = {Radha, Mustafa and Fonseca, Pedro and Moreau, Arnaud and Ross, Marco and Cerny, Andreas and Anderer, Peter and Long, Xi and Aarts, Ronald M.},
  year = 2021,
  journal = {npj Digital Medicine},
  volume = {4},
  number = {1},
  pages = {135},
  doi = {10.1038/s41746-021-00510-8}
}

@book{AASMManualScoringTroester23,
  title = {The {{AASM}} Manual for the Scoring of Sleep and Associated Events: Rules, Terminology and Technical Specifications},
  author = {Troester, Matthew M. and Quan, Stuart F. and Berry, Richard B. and Medicine, American Academy of Sleep},
  year = 2023
}

@article{ScoringVariabilityPolysomnographyCollop02,
  title = {Scoring Variability between Polysomnography Technologists in Different Sleep Laboratories},
  author = {Collop, Nancy A.},
  year = 2002,
  journal = {Sleep Medicine},
  volume = {3},
  number = {1},
  pages = {43--47},
  doi = {10.1016/S1389-9457(01)00115-0}
}

@article{SleepStagingModelZhang25,
  title = {A Sleep Staging Model Based on Adversarial Domain Generalized Residual Attention Network},
  author = {Zhang, Pengwei and Xiang, Sijia and Hu, Kailei and He, Jialing and Chen, Jingxia},
  year = 2025,
  journal = {Frontiers in Neuroscience},
  volume = {19},
  pages = {1501511},
  doi = {10.3389/fnins.2025.1501511},
  pmcid = {PMC12098520},
  pmid = {40415890}
}

@article{SystematicReviewSleepWara25,
  title = {A Systematic Review on Sleep Stage Classification and Sleep Disorder Detection Using Artificial Intelligence},
  author = {Wara, Tayab Uddin and Fahad, Ababil Hossain and Das, Adri Shankar and Shawon, Md Mehedi Hasan},
  year = 2025,
  journal = {Heliyon},
  volume = {11},
  number = {12},
  doi = {10.1016/j.heliyon.2025.e43576}
}

@article{UnsupervisedSleepStagingZhao21,
  title = {Unsupervised Sleep Staging System Based on Domain Adaptation},
  author = {Zhao, Ranqi and Xia, Yi and Zhang, Yongliang},
  year = 2021,
  journal = {Biomedical Signal Processing and Control},
  volume = {69},
  pages = {102937},
  doi = {10.1016/j.bspc.2021.102937}
}

@article{TransferringStructuredKnowledgeYoo22,
  title = {Transferring Structured Knowledge in Unsupervised Domain Adaptation of a Sleep Staging Network},
  author = {Yoo, Chaehwa and Lee, Hyang Woon and Kang, Je-Won},
  year = 2022,
  journal = {IEEE Journal of Biomedical and Health Informatics},
  volume = {26},
  number = {3},
  pages = {1273--1284},
  doi = {10.1109/JBHI.2021.3103614}
}

@article{DACADDomainAdaptationDarban25,
  title = {{{DACAD}}: {{Domain}} Adaptation Contrastive Learning for Anomaly Detection in Multivariate Time Series},
  author = {Darban, Zahra Zamanzadeh and Yang, Yiyuan and Webb, Geoffrey I. and Aggarwal, Charu C. and Wen, Qingsong and Pan, Shirui and Salehi, Mahsa},
  year = 2025,
  journal = {IEEE Transactions on Knowledge and Data Engineering},
  volume = {37},
  number = {8},
  pages = {4485--4496},
  doi = {10.1109/tkde.2025.3569909}
}

@inproceedings{AttentiveAdversarialNetworkNasiri20,
  title = {Attentive Adversarial Network for Large-Scale Sleep Staging},
  booktitle = {Proceedings of the 5th {{Machine Learning}} for {{Healthcare Conference}}},
  author = {Nasiri, Samaneh and Clifford, Gari D.},
  year = 2020,
  pages = {457--478}
}

@article{MoreAccurateAutomaticPhan21,
  title = {Towards More Accurate Automatic Sleep Staging via Deep Transfer Learning},
  author = {Phan, Huy and Ch{\'e}n, Oliver Y. and Koch, Philipp and Lu, Zongqing and McLoughlin, Ian and Mertins, Alfred and De Vos, Maarten},
  year = 2021,
  journal = {IEEE Transactions on Biomedical Engineering},
  volume = {68},
  number = {6},
  pages = {1787--1798},
  doi = {10.1109/TBME.2020.3020381}
}

@article{UnsupervisedDomainAdaptationFan22,
  title = {Unsupervised Domain Adaptation by Statistics Alignment for Deep Sleep Staging Networks},
  author = {Fan, Jiahao and Zhu, Hangyu and Jiang, Xinyu and Meng, Long and Chen, Chen and Fu, Cong and Yu, Huan and Dai, Chenyun and Chen, Wei},
  year = 2022,
  journal = {IEEE Transactions on Neural Systems and Rehabilitation Engineering},
  volume = {30},
  pages = {205--216},
  doi = {10.1109/TNSRE.2022.3144169}
}

@inproceedings{SelfsupervisedContrastiveLearningJiang21,
  title = {Self-Supervised Contrastive Learning for {{EEG-based}} Sleep Staging},
  booktitle = {International {{Joint Conference}} on {{Neural Networks}}},
  author = {Jiang, Xue and Zhao, Jianhui and Du, Bo and Yuan, Zhiyong},
  year = 2021,
  eprint = {2109.07839},
  primaryclass = {cs},
  pages = {1--8},
  doi = {10.1109/ijcnn52387.2021.9533305}
}

@article{ProgressiveInvariantCausalWang25,
  title = {Progressive Invariant Causal Feature Learning for Single Domain Generalization},
  author = {Wang, Yuxuan and Yang, Muli and Wu, Aming and Deng, Cheng},
  year = 2025,
  journal = {IEEE Transactions on Image Processing},
  volume = {34},
  pages = {2694--2706},
  doi = {10.1109/tip.2025.3563772}
}

@inproceedings{LearningDiversifySingleWang21,
  title = {Learning to Diversify for Single Domain Generalization},
  booktitle = {2021 {{IEEE}}/{{CVF International Conference}} on {{Computer Vision}}},
  author = {Wang, Zijian and Luo, Yadan and Qiu, Ruihong and Huang, Zi and Baktashmotlagh, Mahsa},
  year = 2021,
  pages = {814--823},
  doi = {10.1109/iccv48922.2021.00087}
}

@article{ExploringStructureIncentiveMa24,
  title = {Exploring Structure Incentive Domain Adversarial Learning for Generalizable Sleep Stage Classification},
  author = {Ma, Shuo and Zhang, Yingwei and Chen, Yiqiang and Xie, Tao and Song, Shuchao and Jia, Ziyu},
  year = 2024,
  journal = {ACM Transactions on Intelligent Systems and Technology},
  volume = {15},
  number = {1},
  pages = {1--30},
  doi = {10.1145/3625238}
}

@article{DomainInvariantRepresentationLearningLee25,
  title = {Domain-{{Invariant Representation Learning}} and {{Sleep Dynamics Modeling}} for {{Automatic Sleep Staging}}},
  author = {Lee, Seungyeon and Pham, Thai-Hoang and Cheng, Zhao and Zhang, Ping},
  year = 2025,
  journal = {ACM Transactions on Computing for Healthcare},
  volume = {6},
  number = {4},
  eprint = {2312.03196},
  primaryclass = {cs},
  pages = {1--20},
  doi = {10.1145/3757066}
}

@article{MultiViewSpatialTemporalGraphJia21,
  title = {Multi-View Spatial-Temporal Graph Convolutional Networks with Domain Generalization for Sleep Stage Classification},
  author = {Jia, Ziyu and Lin, Youfang and Wang, Jing and Ning, Xiaojun and He, Yuanlai and Zhou, Ronghao and Zhou, Yuhan and Lehman, Li-wei H.},
  year = 2021,
  journal = {IEEE Transactions on Neural Systems and Rehabilitation Engineering},
  volume = {29},
  pages = {1977--1986},
  doi = {10.1109/TNSRE.2021.3110665}
}

@article{WNSleepModelingWholeNightZhou25,
  title = {{{WN-sleep}}: Modeling Whole-Night Data for Improved Sleep Staging Classification},
  author = {Zhou, Fang and Lu, Zhi and Wu, Zhi and Ye, Gaohan and Shu, Lingjie and Pu, Yu and Wang, Beilei and Zhang, Dong and Zhang, Dongheng and Hu, Yang and Chen, Yan},
  year = 2025,
  journal = {IEEE Journal of Biomedical and Health Informatics},
  pages = {0--14},
  doi = {10.1109/JBHI.2025.3613485},
  pmid = {40991602}
}

@inproceedings{PracticalSingleDomainYang24,
  title = {Practical Single Domain Generalization via Training-Time and Test-Time Learning},
  booktitle = {Proceedings of the 30th {{ACM SIGKDD Conference}} on {{Knowledge Discovery}} and {{Data Mining}}},
  author = {Yang, Shuai and Zhang, Zhen and Gu, Lichuan},
  year = 2024,
  series = {{{KDD}} '24},
  pages = {3794--3805},
  doi = {10.1145/3637528.3671806}
}

@inproceedings{ProgressiveDomainExpansionLi21,
  title = {Progressive Domain Expansion Network for Single Domain Generalization},
  booktitle = {2021 {{IEEE}}/{{CVF Conference}} on {{Computer Vision}} and {{Pattern Recognition}} ({{CVPR}})},
  author = {Li, Lei and Gao, Ke and Cao, Juan and Huang, Ziyao and Weng, Yepeng and Mi, Xiaoyue and Yu, Zhengze and Li, Xiaoya and Xia, Boyang},
  year = 2021,
  pages = {224--233},
  doi = {10.1109/CVPR46437.2021.00029}
}

@inproceedings{DomainRandomizationPyramidYue19,
  title = {Domain Randomization and Pyramid Consistency: {{Simulation-to-real}} Generalization without Accessing Target Domain Data},
  booktitle = {{{IEEE}}/{{CVF International Conference}} on {{Computer Vision}}},
  author = {Yue, Xiangyu and Zhang, Yang and Zhao, Sicheng and {Sangiovanni-Vincentelli}, Alberto and Keutzer, Kurt and Gong, Boqing},
  year = 2019,
  pages = {2100--2110},
  doi = {10.1109/ICCV.2019.00219}
}

@inproceedings{ProgressiveRandomConvolutionsChoi23,
  title = {Progressive Random Convolutions for Single Domain Generalization},
  booktitle = {2023 {{IEEE}}/{{CVF Conference}} on {{Computer Vision}} and {{Pattern Recognition}} ({{CVPR}})},
  author = {Choi, Seokeon and Das, Debasmit and Choi, Sungha and Yang, Seunghan and Park, Hyunsin and Yun, Sungrack},
  year = 2023,
  pages = {10312--10322},
  doi = {10.1109/CVPR52729.2023.00994}
}

@inproceedings{GeneralizingUnseenDomainsVolpi18,
  title = {Generalizing to Unseen Domains via Adversarial Data Augmentation},
  booktitle = {{{NeurIPS}}},
  author = {Volpi, Riccardo and Namkoong, Hongseok and Sener, Ozan and Duchi, John C. and Murino, Vittorio and Savarese, Silvio},
  editor = {Bengio, Samy and Wallach, Hanna M. and Larochelle, Hugo and Grauman, Kristen and {Cesa-Bianchi}, Nicol{\`o} and Garnett, Roman},
  year = 2018,
  eprint = {1805.12018},
  primaryclass = {cs},
  pages = {5339--5349}
}

@inproceedings{RobustGeneralizableVisualXu21,
  title = {Robust and Generalizable Visual Representation Learning via Random Convolutions},
  booktitle = {International {{Conference}} on {{Learning Representations}}},
  author = {Xu, Zhenlin and Liu, Deyi and Yang, Junlin and Raffel, Colin and Niethammer, Marc},
  year = 2021,
  eprint = {2007.13003},
  primaryclass = {cs}
}

@inproceedings{ImageNettrainedCNNsAreGeirhos19,
  title = {{{ImageNet-trained CNNs}} Are Biased towards Texture; Increasing Shape Bias Improves Accuracy and Robustness},
  booktitle = {International {{Conference}} on {{Learning Representations}}},
  author = {Geirhos, Robert and Rubisch, Patricia and Michaelis, Claudio and Bethge, Matthias and Wichmann, Felix A. and Brendel, Wieland},
  year = 2019,
  eprint = {1811.12231},
  primaryclass = {cs}
}

@inproceedings{ReducingDomainGapNam21,
  title = {Reducing Domain Gap by Reducing Style Bias},
  booktitle = {2021 {{IEEE}}/{{CVF Conference}} on {{Computer Vision}} and {{Pattern Recognition}} ({{CVPR}})},
  author = {Nam, Hyeonseob and Lee, HyunJae and Park, Jongchan and Yoon, Wonjun and Yoo, Donggeun},
  year = 2021,
  pages = {8686--8695},
  doi = {10.1109/CVPR46437.2021.00858}
}

@inproceedings{AdvSTRevisitingDataZheng24,
  title = {{{AdvST}}: {{Revisiting}} Data Augmentations for Single Domain Generalization},
  booktitle = {Thirty-{{Eighth AAAI Conference}} on {{Artificial Intelligence}}, {{AAAI}} 2024, {{Thirty-Sixth Conference}} on {{Innovative Applications}} of {{Artificial Intelligence}}, {{IAAI}} 2024, {{Fourteenth Symposium}} on {{Educational Advances}} in {{Artificial Intelligence}}, {{EAAI}} 2014, {{February}} 20-27, 2024, {{Vancouver}}, {{Canada}}},
  author = {Zheng, Guangtao and Huai, Mengdi and Zhang, Aidong},
  editor = {Wooldridge, Michael J. and Dy, Jennifer G. and Natarajan, Sriraam},
  year = 2024,
  eprint = {2312.12720},
  primaryclass = {cs},
  pages = {21832--21840},
  doi = {10.1609/AAAI.V38I19.30184}
}

@inproceedings{FriendlySharpnessAwareMinimizationLi24,
  title = {Friendly {{Sharpness-Aware Minimization}}},
  booktitle = {2024 {{IEEE}}/{{CVF Conference}} on {{Computer Vision}} and {{Pattern Recognition}} ({{CVPR}})},
  author = {Li, Tao and Zhou, Pan and He, Zhengbao and Cheng, Xinwen and Huang, Xiaolin},
  year = 2024,
  eprint = {2403.12350},
  primaryclass = {cs},
  pages = {5631--5640},
  doi = {10.1109/CVPR52733.2024.00538}
}

@article{MultiviewMultimodalSystemTorres18,
  title = {A Multiview Multimodal System for Monitoring Patient Sleep},
  author = {Torres, Carlos and Fried, Jeffrey C. and Rose, Kenneth and Manjunath, B. S.},
  year = 2018,
  journal = {IEEE Transactions on Multimedia},
  volume = {20},
  number = {11},
  pages = {3057--3068},
  doi = {10.1109/TMM.2018.2829162}
}

\newpage

\end{document}